\documentclass{article}
\usepackage{amsmath}
\numberwithin{equation}{subsection}
\usepackage{amssymb}
\usepackage[margin=1in]{geometry}
\usepackage{fancyhdr}
\usepackage{graphicx}
\usepackage[justification=centering]{caption}
\usepackage{subcaption}
\usepackage{tikz}
\usepackage{enumitem}
\usepackage{listings}
\usepackage{mathtools}
\usepackage{xcolor}
\usepackage{setspace}
\usepackage{changepage}
\usepackage{hhline}

\newcommand{\sgn}{\text{sgn}}

\newcommand{\subtitle}[1]{\noindent\textbf{#1}\normalsize\\}
\newcommand{\nl}{~\newline}

\usepackage{hyperref}
\hypersetup{colorlinks, 
	citecolor=black,
	linkcolor=black,
	urlcolor=black}
\begin{document}
	\begin{center}
		\noindent\Huge Calibration and Filtering of Exponential L\'{e}vy Option Pricing Models\\
		\vspace{0.06\textheight}
		\large \textsc{Stavros J. Sioutis}$^{(1)}$\\
	\end{center}
	\normalsize
	\vspace{0.05\textheight}
	\begin{adjustwidth}{2cm}{2cm}
	The accuracy of least squares calibration using option premiums and particle filtering of price data to find model parameters is determined. Derivative models using exponential L\'{e}vy processes are calibrated using regularized weighted least squares with respect to the minimal entropy martingale measure. Sequential importance resampling is used for the Bayesian inference problem of time series parameter estimation with proposal distribution determined using extended Kalman filter. The algorithms converge to their respective global optima using a highly parallelizable statistical optimization approach using a grid of initial positions. Each of these methods should produce the same parameters. We investigate this assertion.
	\end{adjustwidth}
	\nl
	\begin{center}
		\includegraphics[scale=0.2]{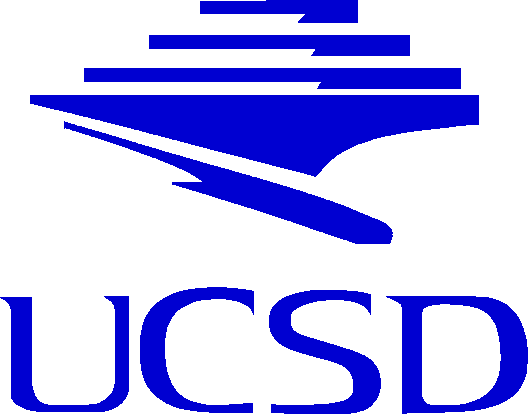}\\
		\large $^{(1)}$Math Department\\University of California, San Diego\\ssioutis@uw.edu
	\end{center}
%	\newpage
%	\tableofcontents
	\section{Introduction}
	This paper presents the most common methods of fitting stock price observations to the mathematical models which capture their stochastic behavior. We first give an introduction to quantitative finance and the necessity of having a fast, accurate method of pricing derivatives. Multiple processes of Exponential L\'{e}vy form are presented. These include both jumps and diffusion components and thus capture a wider range of empirical asset behavior. These generalized processes are then used to price vanilla path-independent options in sections 3 and 4. We show how to empirically fit these models to observed price data. The model fitting methodologies discussed in sections 5 and 6 are backtested to determine their accuracy. These methods are two sides of the same coin, they both retrieve the parameters for a given model which best describe current market conditions, but through complementary means. As such, their predicted parameters should agree. This notion is investigated in section 7.
	
	Arguably the father of the quantitative theory of finance, Harry Markowitz, in his 1952 PhD thesis \textit{Portfolio Selection} described the basis of an entirely new scientific discipline. He developed the concept of treating investment management as an optimization problem based on the mean and variance of each constituent stock. He argued that investors, given their own personal aversion to risk, should hold only those portfolios whose variance is minimized against a specified mean rate of return. 
	
	Stochastic calculus was introduced by Robert Merton in 1969 to accurately price financial securities. This laid the groundwork for Fischer Black and Myron Scholes to develop their famous Black-Scholes-Merton option pricing formula which won the 1997 Nobel prize in Economics. This formula provided a solution to a widely recognized problem: finding the fair market price of a European call option. An option is the right to buy one share of a given stock at a specified price and time.
	
	\subsection{Option pricing in the binomial model}
	
	Consider a stock XYZ whose price is given by a stochastic process $S(t)$, where $S_0$ denotes the price of $S(t) $ at time $ t=0$. Set the current price of $S_0 = \$30$ in a one-period binomial model such that the space of all possible events is defined as $\Omega = \{\omega_u, \omega_d\}$. In the binomial model, the stock price has two possible trajectories, it may go up or down. Let $S_t(\omega_u)$ denote the price of the stock at time $t$ after going up, and $S_t(\omega_d)$ be the price after going down. Here we consider the factors $u, d$ which denote the change in the stock price given the events that took place. We let $u=2$ and $d=1/2$ so that the stock doubles during event $\omega_u$ and halves during event $\omega_d$. The binomial model of the stock price from time $t=0$ to $t=1$ follows
	\begin{center}
		\tikz [ grow=right, level 1/.style={sibling distance=2em}, level distance=4cm]
		\node {$S_0 = 30$} %root
		child {node {$S_1(\omega_d) = S_0\cdot d = 30 / 2 = 15$.}}
		child {node {$S_1(\omega_u) = S_0\cdot u = 30 \cdot 2 = 60$}};
	\end{center}
	Consider a portfolio $X(t)$ with $\Delta$ shares of $S(t)$ and a money market account $M(t) = M_0\cdot (1+r)^t $. $M(t)$ grows at a constant interest rate $r$ with initial capital $M_0$. At time $t=1$ the value of the portfolio $X(t)$ is
	\begin{center}
		\tikz [grow=right, level 1/.style={sibling distance=2em}, level distance=5cm]
		\node{$X(0) = \Delta S_0 + M(0)$} %root
		child {node {$X_1(\omega_d) = \Delta S_1(\omega_d) + M(1)$. }}
		child {node {$X_1(\omega_u) = \Delta S_1(\omega_u) + M(1)$ }};
	\end{center}
	An option is a type of derivative, or contingent claim, which is a financial instrument that \textit{derives} its price from an underlying stock, referred to as the \textit{underlier}. A European call is a type of option that gives the buyer the right but not the obligation (i.e. the option) to buy the stock at an expiration time $T$ and for a given price $K$. We refer to $K$ as the \textit{strike price}. We will price a European call option based on the stock described above.	Our example uses expiration time $T = 1$ and strike price $K = \$40$.
	
	The \textit{terminal condition} of the derivative is its value at expiration, where time $t=T$. If the stock price $S(t)$ is greater than strike price $K$, then the call option value $c(t)$ is equal to $c(T) = (S(T) - K)$, where $S(T)$ is the value of the stock when the call expires. In our example it will be either $S_0\cdot u = \$60$ or $S_0\cdot d = \$15$. If the stock price is below the strike price $K$, then buying the stock using the call option contract would be more expensive than buying the stock on the market, so the option is worthless and we set $c(T) = 0$. Combing the two equations yields $c(T) = \max\{ S(T) - K, 0 \} := (S(T) - K)^+$. Our goal is to price the derivative $c(t)$ at time 0. Our option at the time of expiration $t=T=1$ follows the model
	\begin{center}
		\tikz [ grow=right, level 1/.style={sibling distance=2em}, level distance=6cm]
		\node {$c_0$} %root
		child {node {$(S_T(\omega_d) - K)^+ = (S_1(\omega_d) - K)^+ = (15 - 40)^+ = 0$.}}
		child {node {$(S_T(\omega_u) - K)^+ = (S_1(\omega_u) - K)^+ = (60 - 40)^+ = 20$}};
	\end{center}
	To determine the price of any derivative security, we replicate its payoff. We determine a specific number of shares of $S(t)$ and value of our money market account $M(t)$ which perfectly replicates the value of the derivative $c(t)$ at any time. This is called the \textit{replicating portfolio} and is exactly the portfolio $X(t)$ we developed earlier. Using the value at time $t=1$ of $X(t)$ presents the following system of equations:

	\[c_1(\omega_d) = \Delta S_1(\omega_d) + M(1) = \Delta S_0 \cdot d + M_0\cdot (1+r)^1\]
	\[ c_1(\omega_u) = \Delta S_1(\omega_u) + M(1) = \Delta S_0 \cdot u + M_0\cdot (1+r)^1.\]
	Solving for $\Delta$ and $M_0$ yields
	\[\Delta = \frac{c_1(\omega_u) - c_1(\omega_d)}{S_0\cdot(u-d)}\quad\mbox{and}\quad M_0 = \frac{u\cdot c_1(\omega_u) - d\cdot c_1(\omega_d)}{(1+r)(u-d)}.\]
	The constant $\Delta$ is the \textit{delta hedge} which is the number of shares of $S(t)$ one must hold which, along with $M_0$, will hedge against the risk of holding the derivative. In conclusion, given the delta hedge at time $t=0$, we can determine the value of the replicating portfolio $X(0) = \Delta S_0 + M_0$. This is equal to the value of the derivative at time 0:
	\begin{align*}
	c_0 = X(0) &= \Delta S_0 + M_0\\
	&= \frac{c_1(\omega_u) - c_1(\omega_d)}{S_0\cdot(u-d)} S_0 + \frac{u\cdot c_1(\omega_u) - d\cdot c_1(\omega_d)}{(1+r)(u-d)}\\
	&= \frac{1}{1+r}\bigg[ \frac{1+r-d}{u-d}c_1(\omega_u) + \frac{u-1-r}{u-d}c_1(\omega_d) \bigg]\\
	&= \frac{1}{1+r}\bigg[ \widetilde{p}_uc_1(\omega_u) + \widetilde{p}_dc_1(\omega_d) \bigg].
	\end{align*}
	The vectors $\displaystyle \widetilde{p}_d = \frac{u-1-r}{u-d} $ and $\displaystyle \widetilde{p}_u = \frac{1+r-d}{u-d} $ are \textit{risk neutral probabilities} satisfying $\widetilde{p}_d + \widetilde{p}_u = 1$ and $\widetilde{p}_u,\; \widetilde{p}_d \in \mathbb{R}^\ge$. Here we see that the price of a derivative security at time $t=0$ is a weighted combination of replicating portfolios representing every possible state price at time $t=1$. Taking the limit of this is exactly the motivation behind the continuous time application of stochastic calculus for solving this problem, as we will see later. Given our example with $u=2, d= 1/2 $ and $ R = 1+r = 1$, the risk neutral probabilities are
	\[  \widetilde{p}_u = \frac{2 - 1}{2 - 1/2} = \frac{2}{3} \qquad \text{and} \qquad \widetilde{p}_d = \frac{1-1/2}{2-1/2} = \frac{1}{3} = 1 - \widetilde{p}_u. \]
	We can solve for the initial price of the derivative,
	\[  c(0) = \frac{1}{1+r}\bigg[ \widetilde{p}_uc_1(\omega_u) + \widetilde{p}_dc_1(\omega_d) \bigg]
	= \frac{2}{3}20 + \frac{1}{3}0 = \frac{40}{3} \approx \$13.33. \]
	$\$13.33$ is thus the fair market price of the European call with strike price $\$40$ on a stock with a current price of $\$30$ which we know for certain will, at the time of expiration, have a value of either $\$60$ or $\$15$. The binomial model is the simplest way to conceptualize the process of option pricing and, as thus, it has serious shortcomings. Our example relied on the following assumptions:
	\begin{enumerate}[noitemsep]
		\item Shares of stock can always be subdivided
		\item The interest rates for investing and borrowing are the same
		\item There is zero bid-ask spread, i.e. the purchase price and selling price of a stock are the same
		\item There are only two possibilities of a stock's value in the subsequent period
	\end{enumerate}
	The first three assumptions are required by the Black-Scholes-Merton formula, which we will describe later. The fourth assumption is required by the binomial model. The application of stochastic calculus in describing the stock price as a geometric Brownian motion will overcome this last assumption.
		%
		%
		%***********************************************************************************************
		%***********************************************************************************************
		%
		%	
	\section{Financial Modelling}
	In many continuous-time models of finance, stocks are represented by geometric Brownian motion given by the stochastic differential equation (SDE) $dS(t) = \alpha S(t)dt + \sigma S(t)dW(t)$ where $W(t)$ is a standard Brownian motion, $\alpha$ and $\sigma$ are the drift and volatility factors, respectively. Similarly to the discrete case explored earlier, the replicating portfolio is defined as $X(t) = S(t)\Delta(t) + M(t)$ with differential
	\begin{align*}
	dX(t) &= \Delta(t)dS(t) + (1+r)(X(t) - \Delta(t)S(t))dt\\
	&= \underbrace{(1+r)X(t)dt}_\text{1} + \underbrace{\Delta(t)(\alpha-r)S(t)dt}_\text{2} + \underbrace{\Delta(t)\sigma S(t)dW(t).}_\text{3}
	\end{align*}
	The numbered portions can be thought of as:\\
	1) Portfolio rate of return\\
	2) The risk premium associated with investing in $S(t)$\\
	3) The volatility of $S(t)$\\
	\nl
	We use this basic concept of replication to describe how derivatives can be evaluated.
	\subsection{Evolution of Option Value}
	
	Given the assumption that the option can be represented in continuous time by some stochastic differential equation, the evolution of its value can be calculated in continuous time quite simply. There are two major principles behind the valuation of these derivatives. The first is the notion that any derivative can be hedged, which is equivalent to replicating its payoff, or terminal condition. This results (and is inferred) from the market model being \textit{complete}. The second is the concept of risk-neutral pricing which asserts that, for some risk-free rate $r$, discounting any derivative by $r$ results in the process satisfying the martingale property. The discounted stochastic process defining the behavior of the derivative can then be evaluated by referring to the \textit{risk-neutral measure}. These concepts are revisited in section 2.6.
	
	Consider a function $c(t,S(t))$ that depends only on time $t$ and the price of the underlying asset $S(t)$ that gives us the value of the option at time $t$. By Ito's formula, its differential is:
	\[dc(t,S(t)) = \bigg[ c_t(t,S(t)) + \alpha S(t)c_x(t,S(t)) + \frac{1}{2}\sigma^2S(t)^2c_{xx}(t,S(t))dW(t)  \bigg].\]
	Equating the discounted portfolio value differential $d(e^{-rt}X(t))$ with the discounted option price differential $d(e^{-rt}c(t,S(t)))$ yields the \textit{Black-Scholes-Merton partial differential equation} (BSM pde)
	\[  c_t(t,S(t)) + rxc_x(t,S(t)) + \frac{1}{2}\sigma^2x^2c_{xx}(t,S(t)) = rx. \]
	Solving the BSM pde with the terminal condition $c(T,S(t)) = (S(T) - K)^+$ and boundary conditions
	\[  c(t,0) = 0 \quad \text{ and }\quad \lim\limits_{x\rightarrow\infty}\big[ c(t,x) - (x-e^{-r(T-t)}K) \big] = 0, \quad \forall t \in [0,T] \]
	yields the Black-Scholes-Merton pricing formula for a European call:
	\[ c(t, S(t)) = S(t)N(d_+(\tau, S(t))) - Ke^{-r\tau}N(d_-(\tau, S(t))) \]
	where $\tau = T-t$ is the time to maturity, $N(y) = \displaystyle \frac{1}{\sqrt{2\pi}}\int_{-\infty}^{y}e^{-\frac{x^2}{2}}dz = \frac{1}{\sqrt{2\pi}}\int_{y}^{-\infty}e^{-\frac{x^2}{2}}dz$ is the standard normal distribution and
	\[  d_\pm(\tau, S(t)) = \frac{1}{\sigma\sqrt{\tau}} \bigg[ \log\frac{S(t)}{K} + \bigg( r \pm \frac{\sigma^2}{2} \bigg)\tau \bigg]. \]
	Intuitively, $N(d_+)$ is the risk-adjusted probability that the option will be exercised and $N(d_-)$ is the factor by which the present value of the exercised option exceeds the current stock price.\\
	The Black-Scholes formula is still widely used within the financial industry. The model's constant volatility assumption is inconsistent with observed market prices, however. The simplest addition to the Black-Scholes model is to let volatility be a function of time and price of the underlier. This gives us the \textit{Local Volatility Model} by Derman and Kani [25]. Here we modify the standard geometric Brownian motion so that volatility is a function of time and $S(t)$ while the drift term is a function of time. This is given by
	\[ dS(t) = \alpha(t)S(t)dt + \sigma(t, S(t))dW(t). \]
	This addition results in the generalized BSM pde which prices derivatives based on the local volatility model
	\[ c_t(t,S(t)) + \alpha(t)S(t)c_t(t,S(t)) + \frac{1}{2}\sigma(t,S(t))^2S(t)^2c_{xx}(t,S(t)) = r(t)c(t,S(t)).   \]

	While incorporating a volatility function that changes over time more closely approximates observed prices, volatility in reality is often unpredictable in nature, and many models incorporate a stochastic element to the calculation of volatility. One extension of the local volatility model incorporating stochastic volatility results in the aptly named Stochastic Volatility Model. The evolution of the underlying asset price is given by
	\begin{align*}
	dS(t) &= \alpha S(t)dt + V(t)S(t)dW_1(t),\\
	dV(t) &= \alpha_v(t, V(t))dt +\sigma(t, V(t))dW_2(t).
	\end{align*}
	The two Brownian motions $W_1(t)$ and $W_2(t)$ are correlated by factor $\rho$. The terms $\alpha_v$ and $\sigma$ are the drift and variance of the volatility itself.\\
	
	The most popular is the Heston stochastic volatility model which is described by the following SDEs:
	\begin{align*}
	dS(t) &= \alpha S(t)dt + \sqrt{V(t)}S(t)dW_1(t),\\
	dV(t) &= \kappa (\theta - V(t))dt + \sigma V(t))dW_2(t),
	\end{align*}
	where $\kappa$ represents the volatility's speed of mean reversion, $\theta$ is the long term variance and $\sigma$ is the volatility of the variance [38].

	What has been presented so far are known as pure diffusion models. They have attempted to capture the mean-reverting behavior of stocks by adding a time-dependent drift component while modifying the volatility function of the underlier to capture the dynamic volatility phenomena we observe empirically. The drawback to these models is they assume a smooth transition between the changes of stock prices. What is actually observed is that stock prices often jump instantaneously. It has been proposed by Geman, Madan, and Yor (2001) that stochastic pricing processes need not have a diffusion component but must incorporate jumps. The use of the Poisson process to capture jump characteristics while removing the element of diffusion results in the Variance Gamma process.
	
	\subsection{Variance Gamma process (VG)}
	
	Developed by Madan, Carr, and Chang [51], the VG model was sought to ascertain consistency with the observation that the local log-price movements of stock prices are long-tailed relative to the normal distribution, but approach normality over time [53,55,54,31,11]. Stock prices have been observed to ignore smooth transitions through time and jump to different prices instantaneously. The importance of producing a process capable of admitting Poisson-type jumps is necessary.	The VG process is of L\'{e}vy type with finite moments which accounts for high activity by allowing an infinite number of jumps in any interval of time. Since VG produces finite variation, it can be represented as the sum of two independent and increasing processes, one representing price increases and the other price decreases. It generalizes Brownian motion and models the dynamics of the log stock price. Lacking a diffusion component it is known as a pure jump process. 
	
	To construct the VG process $X(t;\sigma,\nu,\theta)$ we take a Brownian motion with drift $\theta$ and variance $\sigma$ represented by
	\[    b(t, \sigma, \theta) = \theta t + \sigma W(t),   \]
	where $W(t)$ is a standard Brownian motion, $X(t; \sigma, \nu, \theta)$ can be defined in terms of $b(t, \sigma, \theta)$ as:
	\[ X_{VG}(t; \sigma, \nu, \theta) = b(\gamma(t; 1, \nu), \sigma, \theta). \]
	The time change is given by an independent random variable $\gamma(t; 1, \nu)$ with unit mean, positive variance, and follows a gamma density. The characteristic function of the VG model is
	\[ \mathbb{E}(e^{iuX_t}) = \left(\frac{1}{1-iu\theta\nu + \sigma^2u^2\nu/2}\right)^{\frac{t}{\nu}}, \]
	and the asset price is given by
	\[  \ln S(t) = \ln S(0) + (r-q+\omega)t + X(t; \sigma, \nu, \theta)  \quad \equiv \quad S(t) = S(0)\exp\{ (r-q+\omega)t + X(t; \sigma, \nu, \theta). \}\]
	The term $\omega = \frac{1}{\nu}\ln(1-\theta\nu-\sigma^2\frac{\nu}{2})$ is the martingale correction which ensures risk-neutrality, ie.
	\[ \mathbb{E}S_t = S_0e^{(r-q)t}. \]
	\subsection{Analytical Formula for VG Option Price}
	
	Unlike many models in finance there exists an analytical formula for a European option price following VG law [53].	The price of a European option can be determined simply by evaluating the risk-neutral expectation of the terminal condition,
	\[ C_t = \mathbb{E}^{\mathbb{Q}}[e^{-r(T-t)}(S(T)-K)^+], \]
	where $\mathbb{Q}$ is an equivalent measure under which $e^{-rt}S(t)$ is a martingale. We define the unit-time log characteristic function of VG as
	\[ \varphi(u) = \frac{1}{\nu}\ln\bigg( \frac{1}{1+\nu u^2/2} \bigg). \]
	Defining the VG process $N=\{N(t)=b(G(t)), t\in [0,T] \}$ for Brownian motion $b(t)$ and right-continuous process of independent gamma increments $G(t)$, the change of measure density process has the form
	\[ \lambda(t) = \exp\Bigg(\sum_{s\le t} \alpha(\omega, s) \Delta N_s - \int_{0}^{t}\varphi\bigg(\frac{\alpha(\omega,s)}{i}\bigg)ds\Bigg). \]
	The term $\alpha(\omega,s)$ is chosen so that $\lambda(t)$ is a $\mathbb{Q}$-martingale. The authors of [53] show that $e^{-rt}\lambda(t)S(t)$ is a martingale only if $\alpha(\omega,s)$ satisfies
	\[\mu - r = \varphi(\alpha/i) + \varphi(\sigma/i) + \varphi((\alpha+\sigma)/i). \]
	The change of measure density is $\lambda(t)=e^{\alpha N(t)-\varphi(\alpha/i)t}$. It is sufficient to determine the option price at time $t=0$ with maturity at $t=T$. The asset price under a VG process is defined as
	\[ S(t) = S(0)\exp\bigg(\sigma N(t) + \bigg[r + \bigg(\frac{1}{\nu}\ln\frac{1-\nu(\alpha+\sigma)^2/2}{1-\nu \alpha^2/2} \bigg)t\bigg ]\bigg), \]
	so the option price can now be written as
	\[ C(t) = \mathbb{E}^\mathbb{P}[e^{-rt}\lambda(t)(S(t)-K)^+]. \]
	Noting that for large $t$, the VG log characteristic function takes the form
	\[ \varphi(u) = \frac{t}{\nu}\ln\bigg( \frac{1}{1+\nu u^2/2(t/\nu)} \bigg). \]
	Then, since $N(t)/\sqrt{t}\sim \mathcal{N}(0,t)$ for large $t/\nu$ where $\mathcal{N}$ is the normal cumulative density function, we integrate the following with respect to $\mathcal{N}(0,t)$:
	\[ \bigg[S(0)\exp\bigg(\sigma N(t) + \frac{t}{\nu}\ln\bigg[\frac{1-\nu(\alpha+\sigma)^2/2}{1-\nu \alpha^2/2} \bigg]\bigg) - Ke^{-rt}\bigg ]^+e^{\alpha N(t)+(t/\nu)\ln(1-\nu\alpha^2/2)} \]
	which yields the price of a VG European call:
	\begin{align*}
	C(t) &= S(0)e^{(\alpha+\sigma)^2t/2}(1-\nu (\alpha+\sigma)^2/2)^{t/\nu}\mathcal{N}(d_1)-Ke^{-rt+\alpha^2t/2}(1-\nu\alpha^2/2)^{t\nu}\mathcal{N}(d_2),\\
	d_1 &= \frac{\ln(S_0/K)}{\sigma\sqrt{t}} + \Bigg[\frac{r+(1/\nu)\ln\Big(\frac{1-\nu(\alpha+\sigma)^2/2}{1-\nu\alpha^2/2}\Big)}{\sigma} + (\alpha+\sigma)\Bigg]\sqrt{t},\\
	d_2 &= d_1-\sigma\sqrt{t}.
	\end{align*}
	
	The following Matlab implementation of the VG approximation will be used to determine the accuracy of subsequent numerical techniques:
	\lstset{tabsize=3, language=MatLab, numbers=left, numbersep=4pt, frame=single, commentstyle=\color[rgb]{0,0.6,0}}
	\begin{lstlisting}
	function c = VGCall(S0, K, r, T, theta, sigma, nu)	
	alpha = (-theta/sigma);
	a   = (alpha + sigma)^2;
	num = 1 - nu*a/2;
	den = 1 - nu*alpha^2/2;
	d1 = log(S0/K)/(sigma*sqrt(T)) + ((r + 1/nu*log(num/den))/sigma...
		+ alpha+sigma)*sqrt(T);
	d2 = d1 - sigma*sqrt(T);
	c = S0 * exp(a*T/2) * (1 - nu*a/2)^(T/nu)*normcdf(d1) ...
		- K * exp(-r*T + alpha^2*T/2) * (1-nu*alpha^2/2)^(T/nu)*normcdf(d2);
	end
	\end{lstlisting}
	
	\subsection{Variance Gamma with Stochastic Arrival (VGSA)}
	
	Developed by Carr, Geman, Madan and Yor [16], the VGSA model is a modified VG process which allows for volatility clustering through a mean-reverting time change. Volatility clustering has been shown to be present in many markets and so a generalized process capable of accounting for this feature is favorable. The clustering phenomena is achieved through persistent random time changes, which must be mean-reverting. The typical example of a mean-reverting process is the square root model by Cox-Ingersoll-Ross (CIR) [22]. In order to implement VGSA, we evaluate VG at a continuous stochastic time change given by the integral of the CIR process representing the instantaneous stochastic clock. The mean reversion introduced by the CIR process accounts for clustering, also known as volatility persistence, and because of this we are able to calibrate across both strike and maturity simultaneously, unlike VG [39].

	The CIR process $y(t)$ is defined as the solution to the SDE:
	\[ dy(t) = \kappa(\eta - y(t))dt + \lambda\sqrt{y(t)}dW(t), \]
	where $\eta$ is the rate of time change, $\kappa$ is the rate of mean reversion, $\lambda$ is the time change volatility. Since $y(t)$ represents the instantaneous rate of time change, we integrate to get the actual time change for all $t$:
	\[ Y(t) = \int_{0}^{t}y(u)du. \]
	The characteristic function of the time change $Y(t)$ is
	\[ \mathbb{E}e^{iuY(t)} = \varphi(u,t,y(0),\kappa, \eta, \lambda) = A(u,t,\kappa,\eta,\lambda)e^{B(u,t,\kappa,\lambda)y(0)}, \]
	with
	\begin{align*}
		A(u,t,\kappa,\eta,\lambda) &= \frac{\exp\Big( \frac{\kappa^2\eta t}{\lambda^2} \Big) }{\Big( \cosh(\gamma t/2) + \frac{\kappa}{\gamma}\sinh(\gamma t/2) \Big)^{\frac{2\kappa\eta}{\lambda^2}}},\\
		B(u,t,\kappa,\lambda) &= \frac{2iu}{\kappa + \gamma\coth(\gamma t/2)},\\
		\gamma &= \sqrt{\kappa^2-2\lambda^2iu}.
	\end{align*}
	The VGSA process is defined as:
	\[ Z_{VGSA}(t) = X_{VG}(Y(t); \sigma, \nu, \theta) = b(\gamma(Y(t); 1, \nu), \sigma, \theta).\]
	where $\kappa, \lambda, \sigma, \nu, \theta, \eta$ are the six parameters defining the model. The characteristic function is given by
	\[ \mathbb{E}e^{iuZ_{VGSA}(t)} = \varphi(-i\Psi_{VG}(u), t, \nu^{-1}, \kappa, \eta, \lambda), \]
	where $\Psi_{VG}(u)$ is the log characteristic function of VG at unit time defined as
	\[ \Psi_{VG}(u) = -\frac{1}{\nu}\log(1-iu\nu\theta + \sigma^2\nu u^2/2). \]
	We may now define the asset pricing process at time $t$ as
	\[ S(t) = S(0)\frac{e^{(r-q)t+Z_{VGSA}(t)}}{\mathbb{E}e^{Z_{VGSA}(t)}.} \]
	Since $\mathbb{E}e^{Z_{VGSA}(t)} = \varphi(-i\Psi_{VG}(-i), t, \nu^{-1}, \kappa, \eta, \lambda )$, the characteristic function of the log of the asset price is given by
	\[ \mathbb{E}e^{iu\log S_t} = e^{iu(\log S_0 + (r-q)t)} \cdot \frac{\varphi(-i\Psi_{VG}(u), t, \frac{1}{\nu}, \kappa, \eta, \lambda )}{\varphi(-i\Psi_{VG}(-i), t, \frac{1}{\nu}, \kappa, \eta, \lambda )^{iu}}. \]

	\subsection{CGMY process}
	Developed by Carr, Geman, Madan and Yor, the CGMY model accommodates behaviors represented by pure jumps and pure diffusions by allowing its arrival rates and variations to be either finite or infinite. Adjusting the parameters can model a variety of different behaviors. For instance, $Y<0$ allows for finite activity, $0\le Y \le 1$ allows for infinite activity with finite variation, while $1\le Y < 2$ allows infinite activity with infinite variation. The parameter $C$ can be thought of as the measure of overall activity, while $G$ and $M$ are measures of skewness. The CGMY model is defined by its L\'{e}vy measure
	\[\nu(x) = C \bigg[ \frac{e^{-Gx}}{x^{1+Y}}\mathbb{I}_{x>0} + \frac{e^{-M|x|}}{|x|^{1+Y}}\mathbb{I}_{x<0} \bigg]. \]
	While the model cannot be represented by a single SDE, its characteristic function is
	\[\mathbb{E}[e^{iuX_t}] = \exp\{ Ct\Gamma(-Y)((M-iu)^Y - M^Y + (G+iu)^Y - G^Y)  \} .   \]
	The CGMY model is a special case of the tempered stable process, defined as
	\[  \nu(x) = \frac{c_+e^{-\lambda_+x}}{x^{1+\alpha}} \mathbb{I}_{x>0}  +\frac{c_-e^{-\lambda_-|x|}}{|x|^{1+\alpha}} \mathbb{I}_{x<0} .\]
	The CGMY model is an interesting alternative to VG and VGSA since, while being a pure jump process, it allows very fine control over the intensity of the jumps, and can approximate a diffusion process by allowing an infinite number of jumps within any interval.

	\subsection{Risk-Neutral Pricing}
	In order to illustrate the concept of risk-neutral pricing, we first define Girsanov's theorem and the Martingale representation theorem on a multidimensional Brownian motion [67]. Letting $W(t)$ be a vector of Brownian motions
	\[ W(t) = \big( W_1(t), \dots, W_d(t)\big),\]
	Girsanov's theorem in multiple dimensions states, for a fixed positive time $T$ with adapted stochastic process $\Theta(t) = \big( \Theta_1(t), \dots, \Theta_d(t)\big)$ and defining
	\begin{align*}
		Z(t) &= \exp\left\{ -\int_{0}^\Theta(u)dW(u) - \frac{1}{2}\int_{0}^{t}\|\Theta(u)\|^2du\right\},\\
		\widetilde{W}(t) &= W(t) + \int_{0}^{t}\Theta(u)du,
	\end{align*}
	then $\mathbb{E}Z(T) = 1$ and $\widetilde{W}(t)$ is a $d-$dimensional Brownian motion under probability measure $\widetilde{\mathbb{P}}$ defined by
	\[ \widetilde{\mathbb{P}}(A) = \int_AZ(\omega)d\mathbb{P}(\omega), \qquad \forall A\in \mathcal{F}(t),\]
	where $\mathcal{F}(t)$ is the filtration associated with $W(t)$. The multidimensional Girsanov theorem shows that, given the adapted stochastic process $\Theta(t)$ which is path-dependent (adapted) on the Brownian motions $W(t)$, where the components of $W(t)$ are independent of each other, the Brownian motion $\widetilde{W}(t)$ defined using the dependent stochastic process $\Theta(t)$ is certainly dependent itself. However, under the probability measure $\widetilde{\mathbb{P}}$, $\widetilde{W}(t)$ is independent.
	
	Finally, the Martingale representation theorem states that for a positive time $T$ and a filtration $\mathcal{F}(t)$ associated with a $d-$dimensional Brownian motion $W(t)$, for a martingale process $M(t)$ defined on this filtration under the measure $\mathbb{P}$ of the Brownian motions, there exists an adapted, $d-$dimensional process $\Gamma(u)$ such that
	\[ M(t) = M(0) + \int_{0}^{t}\Gamma(u)\cdot dW(u), \qquad\text{for }0\le t \le T. \]
	Concomitantly, by the results of the multidimensional Girsanov theorem explained previously, for a $\widetilde{\mathbb{P}}$-martingale process $\widetilde{M}(t)$, there is an adapted $d-$dimensional process $\widetilde{\Gamma}(u)$ satisfying
	\[ \widetilde{M}(t) = \widetilde{M}(0) + \int_{0}^{t}\widetilde{\Gamma}(u)\cdot dW(u), \qquad\text{for }0\le t \le T. \]
	
	The impact of the previous theorems show that we can construct a multidimensional market model of $m$ stocks, each with their own respective stochastic differential, based on a vector of Brownian motions $W(t)$, each component of which is correlated by some matrix $\rho_{ij}$ which represents the instantaneous correlations between the Brownian motions, which are continuous martingales. Defining each stock to be some stochastic process labeled $S_i(t), i=1:m$, we may describe them in terms of their relative differentials, which combines the notion of the instantaneous correlation matrix along with the instantaneous standard deviations (or volatility processes) $\sigma_i,\; i=1:m$ as
	\[ \frac{dS_i(t)}{S_i(t)}\cdot \frac{dS_j(t)}{S_j(t)} = \rho_{ij}(t)\sigma_i(t)\sigma_j(t).\]
	Then we may define a discount process
	\[ D(t) = e^{-\int_{0}^{t}R(u)du}, \]
	under which the instantaneous correlations and volatility processes remain unchanged, but the mean rates of return are discounted. The motivation behind the risk-neutral measure $\widetilde{\mathbb{P}}$ is that, a stochastic process under $\widetilde{\mathbb{P}}$ is a martingale, which permits us the use of Ito-calculus. Given some interest rate $r$, the discount process converts a probability measure $\mathbb{P}$ into the risk-neutral measure $\widetilde{\mathbb{P}}$ so that, should a mean rate of return of some stock be equal to $r$, under the risk neutral measure its mean becomes 0, which validates the notion of it achieving the martingale property of constant expectation.
	Before we proceed, we define the precise definition or arbitrage (something for nothing). Given some portfolio process $X(t)$ with $X(0) = 0$ and for some positive time $T$ satisfying
	\[ \mathbb{P}(X(T) \ge 0) =1 \equiv \mathbb{P}(X(T) < 0) = 0 \qquad\text{and}\qquad \mathbb{P}(X(T) > 0) > 0, \]
	then $X(t)$ is an arbitrage. Clearly, $X(t)$ represents some trading strategy that can start with zero capital and at some positive time $T$ we are guaranteed to have not lost money while also having some positive probability of making money. Arbitrage is the notion around which our risk-neutral models fail.
	
	This brings us to the first fundamental theorem of asset pricing. If a market model admits a risk-neutral measure, then it does not admit arbitrage. The proof of which is remarkably simple yet intuitively powerful. If a market model has a risk neutral measure $\widetilde{\mathbb{P}}$ then every discounted portfolio process $D(T)X(T)$ is a $\widetilde{\mathbb{P}}$-martingale $\implies \widetilde{\mathbb{E}}[D(T)X(T)] = 0$. Suppose we have some portfolio (hedging) process $X(t)$ with $X(0) = 0$ which is martingale and satisfies the characteristic of an arbitrage ($\mathbb{P}(X(T) < 0) = 0 $), since $\widetilde{\mathbb{P}} \equiv \mathbb{P}$, we have $\widetilde{\mathbb{P}}(X(T) < 0) = 0 $ also. Since $\widetilde{\mathbb{E}}[D(T)X(T)] = 0$ by the martingale property, this implies that $\widetilde{\mathbb{P}}(X(T) > 0) = 0$ by their equivalence. This also means that $\mathbb{P}(X(T) > 0) = 0$ which is a contradiction, therefore $X(t)$ is not an arbitrage. Moreover, since every portfolio process satisfies $X(0) = 0$, there can never be an arbitrage.
	
	The second part of the fundamental theorem of asset pricing shows that a market is complete if and only if the risk-neutral probability measure is unique. Completeness in this sense refers to the concept that every possible derivative can be hedged.
	\section{Transform Methods}
	
	Efficient methods of numerically evaluating complex financial contracts are required. The Feynman-Kac theorem relates the expectation of a stochastic differential equation governing the behavior of the underlying price process with a analytically calculable PDE. Various methods of evaluating these complex derivative payoff functions fall into three major categories, partial integro-differential equation methods, Monte Carlo simulation methods, and numerical transform techniques. The numerical approximation techniques presented in this chapter are among the many methods in the literature that require a transformation to the Fourier domain. The methods presented in this text are applied specifically to path-independent options but, as numerical integration techniques are a heavily studied topic in finance, there exists methods devoted toward their path-dependent counterparts. Efficient techniques for valuating options with early exercise, such as American options, have been developed [14,4,3,49,60]. The convolution (CONV) method can be used with the fast Fourier transform to achieve almost linear complexity on American and Bermudan options and uses the assumption that the probability density of a process can be seen as a transition density which is then written as a convolution integral [49]. The mechanics involve using the fact that the characteristic function of a convolution is the product of the constituting characteristic functions. Amongst others, the saddlepoint method of Carr and Madan is an alternative designed to price deep out-of-the-money options using a modified Lugannani-Rice saddlepoint approximation [15].
	
	In this section we describe the foundation of deriving the Fourier-Cosine method for pricing path-independent derivatives for which the characteristic function is known. The first major development in this arena was by Carr and Madan [14] where they develop the Fast Fourier Transform (FFT) to valuate options efficiently. Given the characteristic function of the risk neutral density, a simple analytic expression of the Fourier transform of the option value can be developed. In every model used in this paper, the characteristic function is known. Since we focus on models with an explicit L\'{e}vy-Khintchine representation, the characteristic function arises naturally. Besides the VG, VGSA, and CGMY processes described earlier, the class of processes for which the characteristic function is known includes: the process of independent increments (McCulloch 1978); the inverse Gaussian (Barndorff-Nielson 1997); pure diffusion with stochastic volatility (Heston 1993); jump processes with stochastic volatility (Bates 1996); and jump processes with stochastic volatility and stochastic interest rates (Scott 1997). While the FFT method presented a considerable breakthrough in terms of computational speed and a complexity of $\mathcal{O}(N\log_2N)$ (with $N^2$ the integration grid size), it has a few shortcomings. It can only be applied to path-independent derivatives with a European-style payoff with an explicit characteristic function. We first describe the concept of transforming the characteristic function of a derivative to determine the discounted expected value of its risk-neutral density.
	\subsection{Laplace transform}
	The use of the Laplace transform in option pricing is to transform a complex PDE into an ODE that is usually easier to solve. We illustrate the Laplace method as a precursor to the Fourier method for evaluating the characteristic function later. The Laplace transform $F(\gamma)$ of a function $F(\tau)$ is defined as
	\begin{equation}
	F(\gamma) =\mathcal{L}(f(\tau))) = \int_{0}^{\infty}e^{\gamma \tau}f(\tau)d\tau  \label{eq:laplace}
	\end{equation}
	where $\gamma\in\mathbb{C}$ and $f(\tau)$ is any function making the integral finite. Any function not satisfying the previous constraint does not have a Laplace transform. Convergence of the integral is satisfied if $\mathcal{R}(\gamma)>\gamma_0$ where $\gamma_0$ is the \textit{abscissa of convergence}. The transform also satisfies linearity:
	\[\mathcal{L}(af_1(\tau) + bf_2(\tau)) = a\mathcal{L}(t_f(\tau)) + b\mathcal{L}(f(\tau)). \]
	Noting that the Laplace transform of a derivative is given by
	\[ \mathcal{L}(f'(\tau)) = \gamma \mathcal{L}(t(\tau))) - f(0^-),    \]
	when the Laplace transform for a function $f(\tau)$ is known, $f(\tau)$ can be recovered using the Bromwich inversion formula:
	\[ f(\tau) = \mathcal{L}^{-1}(F(\tau)) = \lim\limits_{R\rightarrow\infty}\frac{1}{2\pi i} \int_{a-iR}^{a+iR}F(\gamma)e^{\tau\gamma}d\gamma. \]
	The Laplace inversion is highly sensitive to round-off error, and so it is an ill-conditioned problem (Kwok and Barthez 1989). The standard inversion formula is a contour integral which is not a calculable expression. This can be avoided when the Laplace transform is known in closed form as a complex function. Instead of discretizing the forward Laplace integral, we operate the inversion using the Bromwich contour integral while using the transform's values on the complex plane. Letting the contour be the vertical line $\gamma = a$ then the original function $f(\tau)$ is given by
	\[  f(\tau) = \frac{1}{2\pi i} \int_{a-i\infty}^{a+i\infty}e^{\gamma\tau} F(\gamma)d\gamma,\quad \tau >0. \]
	Letting $\gamma = a+iu$ we obtain,
	\begin{equation}
		f(\tau) = \frac{2e^{a\tau}}{\pi} \int_{0}^{\infty}\mathcal{R}(F(a+iu))\cos(u\tau)du
		= -\frac{2e^{a\tau}}{\pi} \int_{0}^{\infty}\mathcal{I}(F(a+iu))\sin(u\tau)du.
		\label{eq:laplaceinversion}
	\end{equation}
	While there are many numerical integration methods which exploit the structure of the inversion formula \eqref{eq:laplaceinversion}, the commonality and recent attention of the Fourier method in finance is the most relevant.
	\subsection{Fourier Method}
	The Fourier series algorithm is a discretization method of solving \eqref{eq:laplaceinversion} proposed by [26] and relies on the following trapezoidal approximation. With step size $\Delta$,
	\[ f_\Delta(\tau) = \frac{\Delta e^{a\tau}}{\pi}\mathcal{R}\left(F\left(a\right)\right) + \frac{2\Delta e^{a\tau}}{\pi}\sum_{k=1}^{\infty}\mathcal{R}\left(F\left(a+ik\Delta\right)\right)\cos(k\Delta\tau). \]
	Letting $\Delta = \frac{\pi}{2\tau}$ and $a = \frac{A}{2\tau}$ gives the alternating series
	
	\[ f_\Delta(\tau) = \frac{e^{\frac{A}{2}}}{2\tau}\mathcal{R}\left(F\left(\frac{A}{2\tau}\right)\right) + \frac{e^{\frac{A}{2}}}{\tau}\sum_{k=1}^{\infty}(-1)^k\mathcal{R}\left(F\left(\frac{A + 2ki\pi}{2\tau}\right)\right) \]
	which eliminates the cosine term. The choice of $A$ must be made so that $a$ falls to the left of the real part of the singularities of the function $F(\gamma)$. This method is surprisingly effective in the context of which the integrands are periodically oscillating, as the errors tend to cancel. The bound of the discretization error, shown by [1], is
	\[ |f(\tau) - f_\Delta(\tau)| < M\frac{e^{-A}}{1-e^{-A}} \simeq Me^{-A}, \]
	for $f(\tau) < M$. If the real term has a constant sign for all $k$, the Euler accelerating algorithm can be convenient [1]. This consists of summing the first $n$ terms of the series and then taking the weighted average of an additional $m$ terms, ie.
	\[ f_\Delta(\tau) \approx E(\tau; n, m) = \sum_{j=0}^{m} {m\choose j} 2^{-m}s_{n+k}(\tau),\]
	where $s_n(\tau)$ is the partial sum
	\[ s_n(\tau) = \frac{e^{\frac{A}{2}}}{2\tau}\mathcal{R}\left(F\left(\frac{A}{2\tau}\right)\right) + \frac{e^{\frac{A}{2}}}{\tau}\sum_{k=1}^{n+j}(-1)^k\mathcal{R}\left(F\left(\frac{A + 2ki\pi}{2\tau}\right)\right). \]
	\newline
	\subtitle{Pricing a European Call via the Fast Fourier Transform}
	\newline	
	Following the work of Carr and Madan [14], we illustrate the most basic case of the method of Fourier inversion to price a European call option. We first represent the option pricing problem in terms of the log price density, which allows us to use the Fourier method to obtain the premium. For some security $S_t$ with probability density function $f(S_t)$, log price density of the underlier $q(s_t)$ where $s_t = \ln (S_t)$ and $k=\ln K$ is the log strike. We obtain the log characteristic function:
	\[ \varphi(\nu) = \mathbb{E}[e^{i\nu s_t}] = \int_{-\infty}^{\infty} e^{i\nu s_t}p(s_t)dS_t. \]
	The value of a European call $C_T(k)$ with strike $K = e^k$ can be expressed in terms of its risk neutral density
	\[   C \mathbb{E}\big[(S_T - K)^+\big] = C\int_{k}^{\infty}e^{-rt}(e^s - e^k)q(s)ds = C_T(k),\]
	where the subscript $T$ for the underlier process has been dropped for simplicity. The constant $C$ is the discount value through which we convert to the equivalent martingale measure under which we take expectations. In general, to convert to the risk-neutral measure we would let $C=e^{-r(T-t)}$. We must modify this function as it is not square-integrable. We define the square-integrable call price by
	\[  c_T(k) = e^{\alpha K}C_T(k) ,\quad \alpha>0\]
	where $\alpha$ is the damping parameter ensuring convergence of the integral which results in the analytical tractability of the Fourier transform. We redefine the characteristic function of the modified option price $c_T(k)$ as
	\[\Psi_T(\nu) = \int_{-\infty}^{\infty}e^{i\nu k}c_T(K)dk = e^{-rT}\int_{-\infty}^{\infty}q(s)\bigg(\int_{-\infty}^{s}e^{(\alpha+i\nu)k}(e^s-e^k)dk\bigg)ds,  \]
	for $\alpha>0$. We can express the modified characteristic function of the option premium $\Psi_T$ in terms of the log characteristic function of the asset price $\varphi(\nu)$. We develop this expression
	\[ \Psi_T(\nu) = \frac{e^{-rT}\varphi(\nu-(\alpha+1)i)}{(\alpha+i\nu)(\alpha + i\nu + 1)}, \]
	then we can use Fourier inversion to retrieve the actual option premium. Since $C_t(k) \in \mathbb{R}$, its Fourier transform $\Psi_T(\nu)$ has even real part and odd imaginary part, so we can write the call option premium as:
	\[ C_t(K) = \frac{e^{-\alpha k}}{2\pi} \int_{-\infty}^{\infty}e^{-i\nu k}\Psi_T(\nu)d\nu = \frac{e^{-\alpha k}}{\pi} \int_{0}^{\infty}e^{-i\nu k}\Psi_T(\nu)d\nu . \]
	Numerical integration of the above formula can be computed quite easily. First we define an upper bound $B$ for the integration then numerically evaluate it using a quadrature method. Letting $N$ be the number of grid intervals, $\Delta\nu = \frac{B}{N} = \eta$ be the spacing between those intervals and $\nu_j = (j-1)\eta$ be the endpoints for the integration intervals for $j=1:N+1$, we can use the trapezoidal approximation method to obtain:
	\begin{align*}
	C_T(k) 	&\approx \frac{e^{-\alpha k}}{\pi} \int_{0}^{B}e^{-i\nu k}\Psi_T(\nu)d\nu\\
	&\approx \frac{e^{-\alpha k}}{\pi} \Big[e^{-i\nu_1 k}\Psi_T(\nu_1) + 2e^{-i\nu_2 k}\Psi_T(\nu_2) + \dots + 2e^{-i\nu_N k}\Psi_T(\nu_N)\\
	&\qquad +e^{-i\nu_{N+1} k}\Psi_T(\nu_{N+1})\Big]\frac{\eta}{2}.
	\end{align*}
	The terms are exponentially decaying so the elimination of the final term to satisfy FFT form does not significantly affect the accuracy of the approximation. This yields:
	\[ C_T(k) \approx \frac{e^{-\alpha k}}{\pi} \sum_{j=1}^{N}e^{-i\nu_jk}\Psi_T(\nu_j)\frac{\eta}{2}(2-\delta_{j-1}). \]
	For a more accurate approach we illustrate the application of Simpson's rule which incorporates different weights into the summation. This will allow us to increase the accuracy of integration with larger values of $\eta$:
	\[ C_T(k) \approx \frac{e^{-\alpha k}}{\pi} \sum_{j=1}^{N}e^{-i\nu_jk}\Psi_T(\nu_j)\frac{\eta}{3}(3+(-1)^j-\delta_{j-1}). \]
	The term $\delta_{j-1}$ is the Kronecker delta function defined as
	\[ \delta_{i} = \begin{cases}
		0, & \text{if} \;i\ne 0\\
		1, & \text{if} \;i=0.
	\end{cases} \]
	While the above direct integration method provides accurate results, it is far from efficient. The Fast Fourier Transform (FFT) originally developed by (Cooley and Tukey 1965) can efficiently compute the sum:
	\[ w(m) = \sum_{j=1}^{N}e^{ -i\frac{2\pi}{N} (j-1)(m-1)} x(j), \quad \text{for }\; m=1:N. \]
	This can be reduced to $\mathcal{O}(N\ln N)$ multiplications using a divide and conquer algorithm to break down discrete Fourier transforms which greatly accelerates the speed at which we can evaluate option premiums. Converting the option pricing problem to FFT form requires us to create a range of strikes around the strike we are particularly interested in. For example, an at-the-money call with log strike $k$ necessitates the definition of the range of log strikes
	\[ k = \beta + (m-1)\Delta k = \beta + (m-1)\lambda\qquad\text{for }\;m=1:N \]
	where $\beta = \ln X_0 - \frac{\lambda N}{2}$. Thus our log strike of interest falls directly center. We will write the integral of $C_T(K)$ as an application of the above summation. The call premium can be written as
	\begin{align*}
	C_T(k) &\approx \frac{e^{-\alpha k}}{\pi} \sum_{j=1}^{N}e^{-i\nu_j k}\Psi_T(\nu_j)w_j\\
	&= \frac{e^{-\alpha k}}{\pi} \sum_{j=1}^{N}e^{-i\lambda\eta(j-1)(m-1)}e^{-i\beta \nu_j}\Psi_T(\nu_j)w_j.
	\end{align*}
	Having the log characteristic function of the asset $X_t$ we choose $\eta$ and define the grid size as a power of 2, ie. let $N=2^n$, $\lambda = \frac{2\pi}{N\eta}, \nu_j = (j-1)\eta$ and determine $\alpha$. Then vector $x$ is constructed as
	\[ \textbf{x} = \begin{pmatrix}
	x_1\\
	x_2\\
	\vdots\\
	x_N
	\end{pmatrix} = \begin{pmatrix}
	\frac{\eta}{2}\frac{e^{-r(T-t)}}{(\alpha+i\nu_1)(\alpha+i\nu_1+1)}e^{-i(\ln X_0-\frac{\lambda N}{2})\nu_1}\varphi(\nu_1-(\alpha+1)i)\\[0.3em]
	\eta\frac{e^{-r(T-t)}}{(\alpha+i\nu_2)(\alpha+i\nu_2+1)}e^{-i(\ln X_0-\frac{\lambda N}{2})\nu_2}\varphi(\nu_2-(\alpha+1)i)\\[0.3em]
	\vdots \\
	\eta\frac{e^{-r(T-t)}}{(\alpha+i\nu_N)(\alpha+i\nu_N+1)}e^{-i(\ln X_0-\frac{\lambda N}{2})\nu_N}\varphi(\nu_N-(\alpha+1)i)
	\end{pmatrix},\]
	where the discount factor $e^{-r(T-t)}$ is the constant $C$ explained earlier which facilitates a measure change, and could be any constant. We input this vector \textbf{x} into the FFT routine which returns the vector \textbf{y} = fft(\textbf{x}) then the $m$ call prices across the range of strikes $k_m, \;m=1:N$ is
	\[\textbf{y} = \begin{pmatrix}
	C_T(k_1) \\
	C_T(k_2) \\
	\vdots\\
	C_T(k_N)
	\end{pmatrix} = \begin{pmatrix}
	\frac{1}{\pi}e^{-\alpha(\ln X_0-\frac{N}{2}\lambda )}\mathcal{R}(y_1)\\[0.3em]
	\frac{1}{\pi}e^{-\alpha(\ln X_0-\big(\frac{N}{2}-1\big)\lambda )}\mathcal{R}(y_2)\\[0.3em]
	\vdots\\
	\frac{1}{\pi}e^{-\alpha(\ln X_0-\big(\frac{N}{2}-(N-1)\big)\lambda )}\mathcal{R}(y_N)
	\end{pmatrix} . \]
	Before describing the FFT algorithm in detail we explain the necessity of the optimal damping parameter.\\
	\newline
	\subtitle{Optimal $\alpha$}
	\newline
	Our approximation using FFT relied on the calculation of damping parameter $\alpha$ which we now determine optimally. The damping parameter ensures that the call price is $L^1$ integrable which is sufficient for the Fourier transform to exist.
	The characteristic functions of the Black-Scholes and Variance Gamma models have a convenient analytical structure which enables the optimal payoff-independent $\alpha$ to be computed exactly.	For Black-Scholes, the authors of [50] show that the optimal $\alpha$ satisfies
	\[ \alpha^* = \min_{\alpha \in\mathbb{R}}\big[-\alpha k + \ln(\varphi(-(\alpha+1)i)^2) \big] = -\frac{d_+}{\eta\sqrt{\tau}}, \]
	where $\displaystyle d_+ = \frac{1}{\sigma\sqrt{\tau}}\bigg[ \log(\frac{S0}{K}) + (r+\frac{\sigma^2}{\sqrt{2}})\tau  \bigg]$.\\
	\nl
	For the Variance Gamma model, the optimal $\alpha$ satisfies
	\[ \alpha^* = \min_{\alpha\in\mathbb{R}}\big[-\alpha k + \ln(\varphi(-(\alpha+1)i)^2) \big] = -\frac{\theta}{\sigma^2}-1+\frac{\tau}{\nu \widetilde{m}}-\sgn(\widetilde{m})\sqrt{\frac{\theta^2}{\sigma^2} + \frac{2}{\nu\sigma^2}+\frac{\tau^2}{\nu^2\widetilde{m}^2} } \]
	where $\widetilde{m} = f-k-\omega\tau$ is a quantity related to the log-moneyness of the option.
	
	\nl
	
	The code for implementing Fast Fourier Transform on the Black-Scholes model is as follows.
	\lstset{tabsize=3, language=MatLab, numbers=left, numbersep=4pt, frame=single, commentstyle=\color[rgb]{0,0.6,0}}
	\begin{lstlisting}[mathescape]
		S0 = 100;		%Spot
		K = 100;			%Strike
		r = 0.05;		%Risk-free rate
		q = 0.0;			%Dividend rate
		T = 5;
		sigma = 0.3;
		N = 2^9;        %Grid size: power of two 
		uplim = 50;     %upper limit for integration
		eta = uplim/N;  %spacing of psi integrand
		
		K = linspace(5,150,N);
		lnS = log(S0);
		lnK = log(K);
		
		%Optimal damping
		alpha = (-1/(sigma*T*eta))*(log(S0./K) + (r+sigma^2/sqrt(2))*T);
		
		lambda = (2 * pi) / (N * eta);  %spacing for log strikes
		
		%log strikes ranging from [lnS-b,lnS+b] (near the money)
		b = (N * lambda) / 2;
		ku = - b + lambda * (u - 1);
		
		u = 1:N;		
		j = 1:N;
		vj = (j-1) * eta;
		%Fourier transform of the modified call price
		phi = cf_BS(vj-(alpha+1).*1i,S0,r,sigma,T)./...
			(alpha.^2 + alpha - vj.^2 + 1i * (2 * alpha + 1) .* vj);
		
		psi = exp(-r*T) * phi.* exp(1i * vj * (b)) * eta;
		psi = (psi/3).*(3+(-1).^j-((j-1)==0));  					%Simpson's rule
		fft_psi = ones(1,N)*exp(-1i*2*pi/N)*(j-1).*(psi-1);	%Discrete FFT
		cp = real(exp(-alpha.*ku).*fft_psi))/pi;   		    	%call price vector
		
		%Determine strikes near the money
		strikeIdx = floor((lnK + b)/lambda + 1); 
		iset = max(strikeIdx)+1:-1:min(strikeIdx)-1;
		xp = ku(iset);
		yp = cp(iset);
		call_price_fft = real(interp1(xp,yp,lnK));		%Linear interpolation
	\end{lstlisting}
	\subsection{Fourier Cosine method}
	
	The FFT method of the previous section introduced significant advantages toward the problem of evaluating the price of a model with only a known characteristic function. Most notably is its ability to generate a series of option prices across a range of strikes in just a single iteration. Introduced by Fang and Oosterlee [32], the Fourier-Cosine (COS) method provides considerable improvements of FFT. While the amortized complexity $\mathcal{O}(n\log n)$ of FFT is certainly less than the COS method, the superiority of the cosine expansion arises when integrating non oscillating functions since the number of terms can be reduced significantly while maintaining the same level of accuracy. Consequently, empirically, the COS method is faster. Another significant advantage is that the derivation of the cosine expansion is disassociated from the terms dependent on the terminal condition, so that complex payoffs and path-dependent options can be priced. The FFT method described explicitly the Fourier transform of the option premium in terms of the log characteristic function of the asset price, whereas the COS method first represents the log density in terms of the Fourier cosine expansion whose coefficients can then be expressed in terms of the log characteristic of the asset. This can be reduced to an analytically calculable integral. The COS method suffers from similar drawbacks as FFT such as the inability to price highly out-of-the-money options. An interval bound must be explicitly determined to truncate the infinite integral to ensure the expansion has a finite number of terms. Consequently, the accuracy of the premium is highly sensitive to the choice of bounds.
	
	We follow the notation of [32]. The Fourier cosine series expansion of a function $f(\theta): \mathbb{R} \rightarrow [0,\pi]$ is 
	\[ f(\theta) = \frac{1}{2}A_0 + \sum_{k=1}^{\infty}A_k\cos(k\theta) = \overline{\sum}_{k=0}^{\infty}A_k\cos(k\theta), \]
	with Fourier cosine coefficient
	\[ A_k = \frac{2}{\pi}\int_{0}^{\pi}f(\theta)\cos(k\theta)d\theta. \]
	The notation $\overline{\sum}$ indicates that the first term is weighted by one-half. The Fourier cosine series expansion can be obtained for functions defined on any finite interval $[a,b]$ with change of variables mapping $a$ to 0 and $b$ to $\pi$, ie.
	\[ \theta = \frac{x-a}{b-a}\pi \qquad \text{so that} \qquad x=\frac{b-a}{\pi}\theta + a. \]
	The expansion then becomes
	\[ f(x) = \overline{\sum}_{k=0}^{\infty}A_k\cos\bigg(k\frac{x-a}{b-a}\pi\bigg) \quad \text{with}\quad A_k = \frac{2}{b-a} \int_{a}^{b}f(x)\cos\left(k\frac{x-a}{b-a}\pi\right)dx. \]
	\subsection{Cosine Coefficients in Terms of the Characteristic Function}
	We know that for any probability density function $p(x)$, the relevant characteristic function is obtained via the Fourier transform
	\[ \mathbb{E}e^{i\nu x} = \varphi(\nu) = \int_{-\infty}^{\infty} e^{i\nu x}p(x)dx. \]
	By evaluating the characteristic function at $\nu = \frac{k\pi}{b-a}$ we define the truncated integral
	\begin{equation}
		\widehat{\varphi}\left(\frac{k\pi}{b-a}\right) = \int_{a}^{b}\exp\bigg\{  ix\frac{k\pi}{b-a} \bigg\}f(x)dx
	\label{eq:costrunc}
	\end{equation}
	Multiplying \eqref{eq:costrunc} by $e^{i\frac{k\pi a}{b-a}}$ yields
	\begin{align*}
		\widehat{\varphi}\left(\frac{k\pi}{b-a}\right)e^{i\frac{k\pi a}{b-a}} &= e^{i\frac{k\pi a}{b-a}}\int_{a}^{b}\exp\bigg\{  ix\frac{k\pi}{b-a} \bigg\}f(x)dx\\
		&= \int_{a}^{b}\exp\bigg\{ ik\pi\bigg(\frac{x-a}{b-a}\bigg) \bigg\}f(x)dx\\
		&= \int_{a}^{b} \bigg( \cos\left(k\pi\bigg[\frac{x-a}{b-a}\bigg]\right) + i\sin\left(k\pi\bigg[  \frac{x-a}{b-a}  \bigg] \right)  \bigg).
	\end{align*}
	Therefore,
	\begin{equation*}
		\mathcal{R}\bigg[ \widehat{\varphi}\bigg(\frac{k\pi}{b-a}\bigg) e^{i\frac{k\pi a}{b-a}} \bigg] = \int_{a}^{b}\cos\bigg( k\pi\bigg[ \frac{x-a}{b-a} \bigg] \bigg) f(x)dx.
	\end{equation*}
	Choosing limits of integration so that $\widehat{\varphi}(\nu) \approx \varphi(\nu)$ gives the cosine coefficient defined as
	\begin{equation*}
		A_k = \frac{2}{b-a}\mathcal{R}\bigg\{ \widehat{\varphi}(\nu)\bigg( \frac{k\pi}{b-a} e^{-i\frac{ka\pi}{b-a}}   \bigg)  \bigg\}
	\end{equation*}
	so that
	\begin{equation}
		A_k \approx F_k = \frac{2}{b-a}\mathcal{R}\bigg\{ \varphi(\nu)\bigg( \frac{k\pi}{b-a} e^{-i\frac{ka\pi}{b-a}}   \bigg)  \bigg\}.
	\end{equation}
	We can now replace $A_k$ with $F_k$ in the series expansion of $f(x)$ within $[a,b]$ to get
		\[ f_1(x) = \overline{\sum}_{k=0}^{\infty}F_k\cos\bigg( k\pi \frac{x-a}{b-a} \bigg),\]
	which is further truncated to obtain
	\begin{equation}
		f_2(x) = \overline{\sum}_{k=0}^{N-1}F_k\cos\bigg( k\pi \frac{x-a}{b-a} \bigg).
	\end{equation}
	Keeping in mind that the error in $f_2(x)$ results from the error of approximating $A_k$ with $F_k$ and the truncation error of replacing the upper limit of summation by $N-1$.
	We are now able to use the derived COS formula to price a European call option. The option value at time $t$ can be written as
	\[ v_1(x,t) = e^{-r\Delta t}\int_{a}^{b}v(y, T)f(y|x)dy. \]
	We then replace the density function $f(y|x)$ by its cosine expansion in $y$
	\[ f(y|x) = \overline{\sum}_{k=0}^{\infty}A_k\cos\bigg(k\frac{x-a}{b-a}\pi\bigg), \]
	so that the price of the option is now
	\[ v_1(x,t) = e^{-r\Delta t}\int_{a}^{b}v(y, T)\overline{\sum}_{k=0}^{\infty}A_k\cos\bigg(k\frac{x-a}{b-a}\pi\bigg)dy. \]
	Interchanging the summation and integration and inserting the definition
	\[ V_k = \frac{2}{b-a}\int_{a}^{b}v(y,T)\cos\bigg(k\frac{x-a}{b-a}\pi\bigg)dy \]
	yields
	\[ v_1(x,t) = \frac{b-a}{2}e^{-r\Delta t}\overline{\sum}_{k=0}^{\infty}A_kV_k. \]
	We continue with the following approximations. The coefficients rapidly decay as $k\rightarrow\infty$ so,
	\[ v_1(x,t) \approx v_2(x,t) = \frac{b-a}{2}e^{-r\Delta t}\overline{\sum}_{k=0}^{N-1}A_kV_k. \]
	The coefficients can be further approximated by $F_k$ as defined earlier to obtain
	\begin{equation}
		v(x,t) \approx v_3(x,t) = e^{-r\Delta t}\overline{\sum}_{k=0}^{N-1}\mathcal{R}\bigg\{ \varphi\bigg(\frac{k\pi}{b-a}; x\bigg)e^{-ik\pi \frac{a}{b-a}} \bigg\}V_k.
	\end{equation}

	\subsection{Numerical solution to a vanilla European option}

	We first define the following:
	\begin{enumerate}[noitemsep]
		\item[(i)] $S_t$ is the price of the underlying security at time t
		\item[(ii)] $K$ is the strike price
		\item[(iii)] $x=\ln\big(\frac{S_t}{K} \big)$ is the log price of the underlier
		\item[(iv)] $y=\ln\big(\frac{S_T}{K} \big)$ is the log price at expiration
	\end{enumerate}
	The log-asset price payoff for a European option follows
	\[ v(y,T) = [\alpha K(e^y - 1)]^+, \]
	where $\alpha = 1$ for a call and $\alpha = -1$ for a put. We first define the cosine series coefficients for $g(y) = e^y$ on $[c,d]\subset[a,b]$,
	\begin{equation}
		\chi_k(c,d) = \int_{c}^{d}e^y\cos\bigg( k\pi\frac{y-a}{b-a} \bigg)dy
	\end{equation}
	and the cosine series coefficients of $g(y)=1$ on the same interval:
	\begin{equation}
		\psi_k(c,d) = \int_{c}^{d}\cos\bigg( k\pi\frac{y-a}{b-a} \bigg)dy.
	\end{equation}
	The proof of which can be found in [32]. This gives us the analytical expression for a vanilla European call option
	\begin{equation}
		V_k^{call} = \frac{2}{b-a}\int_{a}^{b}K(e^y-1)^+\cos\bigg(k\pi\frac{y-a}{b-a}\bigg)dy = \frac{2}{b-a}K(\chi_k(0,b) - \psi_k(0,b))
	\end{equation}
	and a vanilla European put can be written as
	\begin{equation}
		V_k^{put} = \frac{2}{b-a}\int_{a}^{b}K(1-e^y)^+\cos\bigg(k\pi\frac{y-a}{b-a}\bigg)dy = \frac{2}{b-a}K(-\chi_k(0,b) + \psi_k(0,b)).
	\end{equation}
	The authors of [32] propose the following calculation for the range of integration $[a,b]$ of the COS method:
	\begin{equation}
		[a,b] = \bigg[ c_1 \pm L \sqrt{c_2 + \sqrt{c_4}} \bigg]\qquad\text{with }\; L=10 \text{, elusively.}
	\end{equation}
	The notation $c_n$ denotes the $n$-th cumulant of the log asset price at expiration $X=\ln\big(S_T/K\big)$ defined by the cumulant generating function
	\[ G(w) = \ln \mathbb{E}e^{wX} = \ln(\varphi(-iw)), \]
	with characteristic function $\varphi$. The $n$-th cumulant is the $n$-th derivative of the cumulant generating function evaluated at zero, ie.
	\[ c_n = G^{(n)}(0) = \frac{-i\varphi^{(n)}(0)}{\varphi(0)} \]
	The following table describes all cumulants which will be used in this paper, where $w$ is the martingale correction term satisfying $e^{-wt} = \varphi(-i, t)$. \\
	\begin{center}
	{\renewcommand{\arraystretch}{2}
	\begin{tabular}{| c | l |}
		\hline
		BS	 & $c_1 = (r-q)T$\\
			 & $c_2 = \sigma^2T$\\
			 & $c_4 = 0$\\
			 & $w = 0$\\\hline
		VG   & $c_1 = (\mu + \theta)T$\\
			 & $c_2 = (\sigma^2 + \nu \theta^2)T$\\
			 & $c_4 = 3(\sigma^4\nu + 2\theta^4\nu^3 + 4\sigma^2\theta^2\nu^2)T $\\
			 & $w = \frac{1}{\nu}\ln(1-\theta\nu - \sigma^2\nu /2)$\\\hline
		CGMY & $c_1 = \mu T + CT\Gamma(1-Y)(M_{Y-1}-G^{Y-1})$\\
			 & $c_2 = \sigma^2T + CT\Gamma(2-Y)(M^{Y-2} + G^{Y-2})$\\
			 & $c_4 = CT\Gamma(4-Y)(M^{Y-4} + G^{Y-4})$\\
			 & $w = -C\Gamma(-Y)[(M-1)^Y - M^Y + (G+1)^Y - G^Y]$\\\hline
		VGSA & Compute using finite differences\\\hline
	\end{tabular}}
	\end{center}
	\subsection{Fourier-Cosine Algorithm}

	Here we implement the Fourier Cosine transform to price a call following the Variance Gamma model\\
	\lstset{tabsize=3, language=MatLab, numbers=left, numbersep=4pt, frame=single, commentstyle=\color[rgb]{0,0.6,0}}
	\begin{lstlisting}[mathescape]
	S0 = 100; K = 90; T = 5; r = .1; sigma = 0.3; nu = 0.2; theta = -.14;
	
	K = linspace(10,150,500);
	
	chi = @(k,a,b,c,d) (1./(1+(k*pi/(b-a)).^2).*(cos(k*pi*(d-a)/(b-a))...
		.*exp(d)-cos(k*pi*(c-a)/(b-a))*exp(c)+k*pi/(b-a).*sin(k*pi*(d-a)...
		/(b-a))*exp(d)-k*pi/(b-a).*sin(k*pi*(c-a)/(b-a))*exp(c)));
	
	psi = @(k,a,b,c,d) ( [d-c;(sin(k(2:end)*pi*(d-a)/(b-a))...
		-sin(k(2:end)*pi*(c-a)/(b-a))).*(b-a)./(k(2:end)*pi)] );
	
	%Variance Gamma characteristic function
	cf_VG = @(u,S,K,t,r,sigma,nu,theta)  ((exp(u*(r+log(1-theta*nu-sigma^2...
		*nu/2)/nu)*t*1i)).*((1-1i*theta*nu*u +sigma^2*nu*u.^2/ 2).^(-t/nu)));
		
	N = 2^8;			%Grid size
	
	%Variance Gamma cumulants
	c1 = (r+theta)*T;
	c2 = (sigma^2+nu*theta^2)*T;
	c4 = 3*(sigma^4*nu+2*theta^4*nu^3+4*sigma^2*theta^2*nu^2)*T;

	%Truncation range
	L = 10;
	a = c1-L*sqrt(c2+sqrt(c4));
	b = c1+L*sqrt(c2+sqrt(c4));
	
	%Compute Fourier Cosine transform
	x = log(S0./K);
	k = (0:N-1)';
	Vk = 2/(b-a)*(chi(k,a,b,0,b) - psi(k,a,b,0,b));	%Cosine coefficients
	w = [.5 ones(1,N-1)];							 		%weights
	ret = w*(cf_VG(k*pi/(b-a),S0, K, T, r, sigma, nu, theta)...
		*ones(1,length(K)).*exp(1i*k*pi*(x-a)/(b-a)).*(Vk*ones(1,length(K))));
	%Call price vector
	cp =K*exp(-r*T).*real(ret);
	\end{lstlisting}
	
	\subsection{VGSA Cumulants}
	
	Since we do not have the cumulants for VGSA in closed form, we must compute them using some discretization technique. We showed earlier that the characteristic function of the VGSA model is defined as a specific parameterization of the characteristic function of the CIR time-change:
	\[ \mathbb{E}e^{iuY(t)} = \varphi_{VGSA}(u,t,y(0),\kappa, \eta, \lambda) = A(u,t,\kappa,\eta,\lambda)e^{B(u,t,\kappa,\lambda)y(0)}, \]
	where
	\begin{align*}
	A(u,t,\kappa,\eta,\lambda) &= \frac{\exp\Big( \frac{\kappa^2\eta t}{\lambda^2} \Big) }{\Big( \cosh(\gamma t/2) + \frac{\kappa}{\gamma}\sinh(\gamma t/2) \Big)^{\frac{2\kappa\eta}{\lambda^2}}},\\
	B(u,t,\kappa,\lambda) &= \frac{2iu}{\kappa + \gamma\coth(\gamma t/2)},
	\end{align*}
	and $\gamma = \sqrt{\kappa^2-2\lambda^2iu}$. The VGSA characteristic function is given by
	\[ \mathbb{E}e^{iuZ_{VGSA}(t)} = \varphi_{VGSA}(-i\Psi_{VG}(u), t, \nu^{-1}, \kappa, \eta, \lambda), \]
	where $\Psi_{VG}(u)$ is the log characteristic function of VG at unit time
	\[ \Psi_{VG}(u) = -\frac{1}{\nu}\log(1-iu\nu\theta + \sigma^2\nu u^2/2). \]
	The cumulant generating function satisfies:
	\begin{align*}
		G(w) &= \ln \mathbb{E}e^{wX} = \ln(\varphi_{VGSA}(-iw))
	\end{align*}
	Its $n$-th cumulant is the $n$-th derivative of the cumulant generating function with respect to $u$ evaluated at zero:
	\[ c_n = G^{(n)}(0) \]
	The first four cumulants are the first four finite differences of $G(w)$:
	\begin{align*}
	c_1 &\approx \frac{G(h)-G(-h)}{2h}\\
	c_2 &\approx \frac{G(h)-2G(0)+G(-h)}{h^2}\\
	c_3 &\approx \frac{G(2h)-2G(h)+2G(-h)-G(-2h)}{2h^3}\\
	c_4 &\approx \frac{G(3h)-2G(2h)+4G(0) - G(h) - G(-h) - 2G(-2h) + G(-3h)}{4h^4}
	\end{align*}
	For some small value $h=10^{-6}$.

	\subsection{Numerical Results of Cosine Method}
	Relative error of FFT using the Black-Scholes model with parameters $S_0=100, r=0.1, q=0, \sigma=0.2, N=2^7$, and varying strikes and times to maturity.\\	
	\begin{figure}[h!]
		\centering
		\includegraphics[scale=0.6]{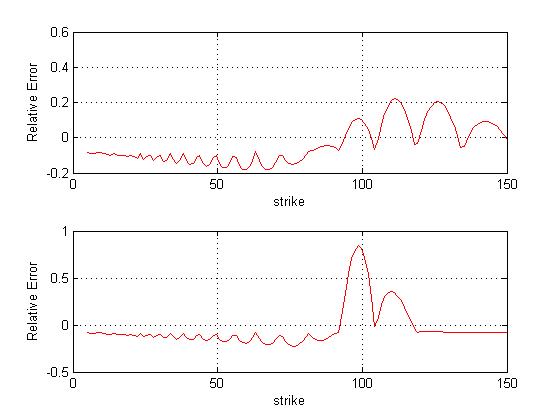}
		\caption{Top: T=1, Bottom, T=0.1}
	\end{figure}
	
	Reference option premium $c = 13.2697$ priced analytically with Black Scholes using parameters $S_0 = 100, K=100,r=0.1,q=0,\sigma =0.2, T=1$\\
	\begin{center}
	\begin{tabular}{|c|c|c|c|c|c|}
		\hline
		\multicolumn{6}{|c|}{COS Method Black-Scholes}\\\hline
		\multicolumn{2}{|c|}{}&\multicolumn{2}{|c|}{T=1}&\multicolumn{2}{|c|}{T=0.1}\\\hline
		K & N & Relative error & CPU time & Relative error & CPU time\\ \hhline{|=|=|=|=|=|=|}
		80 & $2^5$ & -0.0028 & 0.002692 & -1.02e-0.5 & 0.000292\\\hline
		100 & $2^5$ & -1.115e-05 & 0.002148 & -2.88e-6 & 0.000255\\\hline
		120 & $2^5$ & -2.801e-07 & 0.001623 & -2.20e-8 & 0.000275\\\hline	
	\end{tabular}\\
		\end{center}\nl
	Reference option premium priced analytically with Variance Gamma using parameters $S_0 = 100, K=100,r=0.1,q=0,\theta = 0.15\sigma =0.2,\nu=0.1, T=1$.
	\begin{center}
	\begin{tabular}{|c|c|c|c|c|c|c|c|}
		\hline
		\multicolumn{8}{|c|}{COS Method Variance Gamma}\\\hline
		& & \multicolumn{3}{|c|}{T=1}&\multicolumn{3}{|c|}{T=0.1 }\\\hline
		K & N & premium* & Relative error & CPU time & premium* & Relative error & CPU time\\ \hhline{|=|=|=|=|=|=|=|=|}
		& $2^5$ &			 & 0.8730 & 0.001167 &				& 0.1920 & 0.000635\\
		80 & $2^6$ & 28.1547 & 0.1781 & 0.000548 	& 20.8165   & 0.2375 & 0.000564\\
		& $2^7$ &			 & 0.1762 & 0.000593 &				& 0.2429 & 0.000586\\
		&&&&&&&\\
		& $2^5$		&			& 0.00908 & 0.001223 &			& 0.0109 & 0.000675\\
		100 & $2^6$ &13.4251	& 0.0897  & 0.000576 & 3.0543   & 0.0166 & 0.000573\\
		& $2^7$ 	&			& 0.0897  & 0.000586 &			& 0.0165 & 0.000578\\
		&&&&&&&\\
		& $2^5$ 	&& 0.2677 & 0.001294 && 0.0950 & 0.001272\\
		120 & $2^6$ &4.7984& 0.2650 & 0.000592 &0.0068& 0.1012 & 0.000569\\
		& $2^7$ 	&& 0.2650 & 0.000593 && 0.1003 & 0.000580\\
		\hline
	\end{tabular}
		\flushleft\hspace{1cm}*reference premiums determined analytically.
	\end{center}
	\section{Monte Carlo Simulation Methods}
	
	Using Monte Carlo to evaluate the price of an option is conceptually the simplest yet the most computationally intensive method. It does not suffer from the mispricing phenomena of the Fourier Cosine method in pricing out of the money options, yet to get an accurate evaluation one must generally run Monte Carlo over a high number of asset pricing paths.
	
	\subsection{Antithetic Variates}
	When using Gaussian variables to drive a Monte Carlo simulation, we can take advantage of the fact that a standard normal random variable $Z$ has an identical distribution to its reflection $-Z$. The central limit theorem allowing the standard error of Monte Carlo sampling to be determined required independent draws. If we view two random variates $v = Z$ and $\hat{v} = -Z$ as individual samples then we can take the pairwise average of the two $\bar{v} = \frac{1}{2}(v + \hat{v})$ and consider it as an individual sample.
	
	\subsection{Variance Gamma process}

	As explained earlier, the VG process can be obtained from evaluating a Brownian motion at a random time given by the gamma process $\gamma(t;1,\nu)$ (called the subordinator) and follows
	\[ X(t;\sigma,\nu,\theta) = \theta\gamma(t;1,\nu) + \sigma W(\gamma(t;1,\nu)), \]
	with characteristic function
	\[ \varphi(u) = \mathbb{E}e^{iuX_t} = \big( 1-iu\theta\nu + \sigma^2u^2\nu/2 \big)^{-t/\nu}.  \]
	Then the log asset price at time $t$ is given by
	\[ \ln S_t = \ln S_0 + (r-q+\omega) t + X(t;\sigma,\nu,\theta), \]
	where $\omega = \displaystyle -\frac{1}{t}\log(\varphi(-i)) = \frac{1}{\nu}\ln(1-\theta\nu - \sigma^2\nu/2)$ is the martingale correction. We assume $N$ equidistant intervals of length $h = T/N$, then sample from a gamma distribution with mean $k$ and variance $\nu k$. Recall a gamma density function with shape $a$ and scale $b$ follows
	\[   p_\Gamma (x) = \frac{x^{k-1}e^{-\frac{x}{\theta}}}{\theta^k\Gamma(k)}. \]
	Thus, its mean and variance are
	\begin{align*}
		\mu &= ab = h,\\
		\sigma^2 &= ab^2 = \nu h,
	\end{align*}
	which implies the scale and shape are $b = \nu, a = \frac{h}{\nu}$ respectively. So, given random variables $Z \sim N(0,1)$ and $G \sim \Gamma(\frac{h}{\nu}, \nu)$, we obtain a sample for the VG process
	\[X_{VG}(h; \sigma, \nu, \theta) = \theta G + \sigma\sqrt{G}Z. \]
	The following is our Matlab implementation of the Monte Carlo pricing of VG using subordinated Brownian motion.
	\lstset{tabsize=3, language=MatLab, numbers=left, numbersep=4pt, commentstyle=\color[rgb]{0,0.6,0}}
	\begin{lstlisting}[mathescape]
		S = 100; T = 5; r = .1; q = 0; sigma = 0.3; nu = 0.2; theta = -.14;
		
		K = linspace(10,150,50);
		numPaths = 10000;
		N = 100;    %Number of time steps per path
		
		%Martingale correction
		omega = 1/nu*log(1-theta*nu-sigma^2*nu/2);
		h = T/N;	%Time step
		lnS = 0;
		payoff = 0;
		%Gamma time changed Brownian motion
		for j = 1:numPaths
			lnS = log(S);
			for i = 1:N
				Z = normrnd(0,1);
				G = gamrnd(h/nu,nu);
				%Cumulative log asset price
				lnS = lnS + (r-q+omega)*h + theta*G + sigma*sqrt(G)*Z;
			end
			%Terminal condition
			payoff = payoff + max(exp(lnS)-K,0);
		end
		%Discounted average is the price of a European call
		payoff = exp(-r*T)*payoff/numPaths;
	\end{lstlisting}

	\subsection{CGMY tempered stable process}

	Madan and Yor [52] show that the CGMY process can be represented as a time-changed Brownian motion. The CGMY subordinator is absolutely continuous with respect to a one sided stable subordinator achieved by the truncation method proposed by Rosinski. Suppose the CGMY process $X(t)$ is obtained by a subordinated Brownian motion with measure $\nu(dy)$, then by Sato [64] the CGMY L\'{e}vy measure is given by
	\[ \mu(dx) = dx\int_{0}^{\infty}\nu(dy)\frac{1}{\sqrt{2\pi y}}e^{-\frac{(x-\theta y)^2}{2y}}, \]
	where $Y(t)$ is the independent subordinator. The CGMY process can be written as
	\[ X(t) = \theta Y(t) + W(Y(t)). \]
	In particular, the authors of [52] show that the subordinator $Y(t)$ is absolutely continuous with the one sided stable $\frac{Y}{2}$ subordinator with the following L\'{e}vy measure
	\begin{align*}
		\nu(dy) &= \frac{K}{y^{1+\frac{Y}{2}}}f(y)dy,\\
		f(y) &= e^{-\frac{(B^2-A^2)y}{2}}\mathbb{E}\bigg[e^{-\frac{B^2y}{2}\frac{\gamma_{Y/2}}{\gamma_{1/2}}}\bigg],\\
		B &= \frac{G+M}{2}, A = \frac{G-M}{2},\\
		K &= \Bigg[\frac{C\Gamma(\frac{Y}{2})\Gamma(1-\frac{Y}{4})}{2\Gamma(1+\frac{Y}{2})}\Bigg],
	\end{align*}
	with the independent gamma variates $\gamma_{Y/2} \sim \Gamma(Y/2,1)$ and $\gamma_{1/2} \sim \Gamma(1/2,1)$. We write the CGMY subordinator L\'{e}vy measure as
	\begin{align*}
		\nu_1(dy) &= \nu_0(dy)\mathbb{E}\big[e^{-yZ}\big],\\
		\nu_0(dy) &= \frac{K}{y^{1+\frac{Y}{2}}},\\
		Z &= \frac{B^2}{2}\frac{\gamma_{Y/2}}{\gamma_{1/2}}	.
	\end{align*}
	While the expression $\mathbb{E}e^{-yZ}$ can be simulated, it incurs unnecessary randomness, and can be avoided by evaluating explicitly the Laplace transform of $Z$:
	\begin{align*}
		\mathbb{E}\big[e^{-yZ}\big] &= \frac{\Gamma(\frac{Y+1}{2})}{\Gamma(Y)\sqrt{\pi}}2^Y\bigg(\frac{B^2y}{2}\bigg)^{\frac{Y}{2}}I(Y,B^2y, B^2y/2), \\
		I(Y,a,\lambda) &= (2\lambda)^{\frac{Y}{2}}\Gamma(Y)e^{\frac{a^2}{8\lambda}}D_{-Y}\bigg(\frac{a}{\sqrt{2\lambda}}\bigg),
	\end{align*}
	where $D$ is the parabolic cylinder function
	\[ D_p(z) = 2^{\frac{p}{2}}e^{-\frac{z^2}{4}}\Bigg[\frac{\Gamma(\frac{1}{2})}{\Gamma(\frac{1-p}{2})}U\Big(-\frac{p}{2}, \frac{1}{2};\frac{z^2}{2}) - \frac{\sqrt{2\pi}z}{\Gamma(-\frac{p}{2})}U\Big(\frac{1-p}{2}, \frac{3}{2};\frac{z^2}{2}) \Bigg], \]
	and $U$ is the confluent hypergeometric function of the first kind [35]. Since we have identified two L\'{e}vy measures satisfying
	\[ \frac{d\nu_1}{d\nu_0}=\mathbb{E}e^{-yZ} < 1, \]
	it is shown in Rosinski [63] that we may simulate the paths of the CGMY subordinator $\nu_1$ from the paths of $\nu_0$ by rejecting all jumps smaller than a predetermined truncation level $\epsilon$. Letting
	\[ A = \frac{G-M}{2}\qquad\text{and}\qquad B = \frac{G+M}{2}, \]
	we simulate from a one sided stable subordinator with measure:
	\[ \frac{1}{y^{\frac{Y}{2}+1}} \quad \text{and arrival rate}\quad \lambda = \int_{\epsilon}^{\infty}\frac{1}{y^{\frac{Y}{2}+1}}dy = \frac{2}{Y\epsilon^{\frac{Y}{2}}}. \]
	In our implementation, we truncate the jumps below $\epsilon=10^{-4}$ then replace them with their expected drift
	\[ d = \int_{0}^{\epsilon}y\frac{1}{y^{\frac{Y}{2}+1}}dy = \frac{\epsilon^{1-\frac{Y}{2}}}{1-\frac{Y}{2}}. \]
	The jump times are simulated by the exponential random variable
	\[ t_i = -\frac{1}{\lambda}\log(1-U_{1i}). \]
	For uniform sequence of random variables $U_i$ which gives cumulative jump times of
	\[ \Gamma_j = \sum_{i=1}^{j}t_j,\quad\text{and magnitude}\quad y_j = \frac{\epsilon}{(1-U_{2j})^{\frac{2}{Y}}}, \]
	the stable subordinator is given by the process
	\[ S(t) = dt + \sum_{j=1}^{\infty}y_j\mathbb{I}_{\Gamma_j<t},\]
	and the CGMY subordinator is given by
	\[ H(t) = dt + \sum_{j=1}^{\infty}y_j\mathbb{I}_{\Gamma_j<t}\mathbb{I}_{h(y)>U_3}. \]
	For independent uniform sequence $U_3$, with:
	\begin{align*}
		h(y) &= \exp\bigg(-\frac{B^2y}{2}\bigg)\frac{\Gamma(\frac{Y+1}{2})}{\Gamma(y)\sqrt{\pi}}2^Y\bigg(\frac{B^2y}{2}\bigg)^{\frac{Y}{2}}I(Y,B^2y, B^2y/2), \\
		I(Y,a,\lambda) &= (2\lambda)^{\frac{Y}{2}}\Gamma(Y)e^{\frac{a^2}{8\lambda}}D_{-Y}\bigg(\frac{a}{\sqrt{2\lambda}}\bigg).
	\end{align*}
	Then, the CGMY random variable is
	\[ X = AH(t)+\sqrt{H(t)}z, \quad z\sim \mathcal{N}(0,1). \]
	Our algorithm for Monte Carlo simulation of CGMY using subordinated Brownian motion and Rosinski rejection is:
	\lstset{tabsize=3, language=MatLab, numbers=left, numbersep=4pt, frame=single, commentstyle=\color[rgb]{0,0.6,0}}
	\begin{lstlisting}[mathescape]
	%Returns the laplace transform of the CGMY subordinator
	function [ret] = CGMYSubLaplace(y,Y,B)
		ret = (exp((-B^2.*y)/2).*gamma((Y+1)/2)*2.^Y...
			.*(B^2.*y./2).^(Y/2).*I(Y,B^2.*y,B^2.*y./2))/(gamma(Y)*sqrt(pi));
	%Parabolic cylinder function (Laplace change of variables)
	function[ret] = ParCyl(p,z)
		S=1;
		c=1;
		g=gamma(-p/2+0.5);
		H=gamma(-p/2);
		if z<40
			u = Hypergeometric(-p/2,0.5,z.*z./2);
			v = Hypergeometric(0.5-p/2,1.5,z.*z./2);
			D = 2^(p/2).*exp(-z.*z./4).*((sqrt(pi).*u./g)-sqrt(2*pi).*z.*v./H);
		else
			for i=1:20
				c = c * -(p-i+1)*(p-i)./(2*i.*z.^(2*i));
				S = S+c;
			end
			D = exp(-z.^2./4).*z.^p.*S;
		end
		ret = D;
	end
	%Confluent hypergeometric function of the first kind
	function[ret] = Hypergeometric(a,b,z)
		ret=1;
		term=ones(size(z));
		n=1;
		while max(term)>1E-4 && n<100
			term=term.*((a+n-1)*z/(n*(b+n-1)));
			ret=ret+term;
			n=n+1;
		end
	end
	
	function[ret] = I(Y,a,lambda)
		ret = (2.*lambda).^(-Y/2).*gamma(Y).*exp(a.^2./(8.*lambda))...
			.*ParCyl(-Y,(a./sqrt(2.*lambda)));
	end
	end
	
	%Monte Carlo simulation
	C=10; G=10; M=10;
	%Y < 2 higher values means small jumps have greater influence
	Y=0.75;
	numPaths=100;
	T=1;
	S0 = 100; strike = 100;
	St = zeros(1,numPaths);
	r = 0.08;
	payoff = 0;
	A=(G-M)/2;
	B=(G+M)/2;
	epsilon=1E-10; %Jump truncation level	
	d=epsilon^(1-Y/2)/(1-Y/2); %Expected drift of truncated jumps
	lambda=2/(epsilon^(Y/2)*Y); 	%Arrival rate of jumps
	pathSize = ceil(lambda*(T)+1);	%Expected number of jumps before maturity
	warning('off','all');
	tic
	for i =1:numPaths
		%Vector of jump times
		tj=0;
		while tj(end)<T
			U2=rand(1 ,pathSize);
			ti=-log(1-U2)/lambda;
			tj=[tj tj(end)+cumsum(ti)];
		end
		tj=tj(tj<T);    %Reject jumps occuring after maturity
	
		U1=rand(1,length(tj)-1);
		%Vector of jump magnitudes
		yj =[0, epsilon./(1-U1).^(2/Y)];
	
		U3=rand(size(yj));
		%CGMY subordinator, Rosinski rejection
		Ht = d*tj + cumsum(yj.*(CGMYSubLaplace(yj,Y,B)>U3));
		%CGMY random variable
		X=A*Ht+sqrt(Ht).*randn(size(Ht));	
		St(i) = S0*exp(r*T+X(end));
		payoff = payoff + max(St(i)-strike,0);
	end
	toc
	premium = exp(-r*T)*payoff/numPaths
	\end{lstlisting}
	\subsection{Variance Gamma with Stochastic Arrival}

	Recall the CIR process $y(t)$ is the solution to the SDE:
	\[ dy(t) = \kappa(\eta - y(t))dt + \lambda\sqrt{y(t)}dW_t. \]
	The time change is given by its integral,
	\[ Y(t) = \int_{0}^{t}y(s)ds, \]
	which has characteristic function
	\[ \mathbb{E}e^{iuY(t)} = \varphi(u,t,y(0),\kappa,\eta,\lambda) = A(t,u)e^{B(t,u)y(0)}, \]
	where
	\begin{align*}
	A(t,u) &= \frac{\exp{\frac{\kappa^2\eta t}{\lambda^2}}}{\big(\cosh(\gamma t/2) + \frac{\kappa}{\gamma} \sinh(\gamma t/2 )\big)^{2\kappa\eta/\lambda^2}}, \\
	B(t,u) &= \frac{2iu}{\kappa + \gamma \coth(\gamma t/2)},\\
	\gamma &= \sqrt{\kappa^2 - 2\lambda^2iu}.
	\end{align*}
	The VGSA process is defined by the modified VG process
	\[Z(t) = X_{VGSA}(Y(t);\sigma, \nu, \theta) = \theta\gamma(Y(t);1,\nu) + \sigma W(\gamma(Y(t);1,\nu)). \]
	Its characteristic function is that of the time change $Y(t)$ with the following parameters:
	\[ \mathbb{E}e^{iuZ(t)} = \varphi(-i\psi_{VG}(u), t, \frac{1}{\nu}, \kappa, \eta, \lambda ),  \]
	where $\psi_{VG}(u) = \displaystyle - \frac{1}{\nu} \log(1-iu\theta\nu + \sigma^2\nu u^2/2)$ is the log of the characteristic function of the VG process.\\
	The characteristic function of the log of the stock price is:
	\[ \mathbb{E}e^{iu\log S_t} = e^{iu(\log S_0 + (r-q)t)} \cdot \frac{\varphi(-i\psi_{VG}(u), t, \frac{1}{\nu}, \kappa, \eta, \lambda )}{\varphi(-i\psi_{VG}(-i), t, \frac{1}{\nu}, \kappa, \eta, \lambda )^{iu}} \]
	To simulate the VGSA process we again assume $N$ equidistant time intervals of length $h=T/N$. Discretizing its differential gives:
	\begin{align*}
	\Delta Z(t) &= Z(t) - Z(t-h)\\
	&= \theta\gamma(Y(t); 1,\nu) + \sigma W(\gamma(Y(t); 1\nu)) - (\theta \gamma(Y(t-h);1,\nu) + \sigma W(\gamma(Y(t-h);1\nu))\\
	&= \theta(\gamma(Y(t);1,\nu) - \gamma(Y(t-h);1,\nu)) + \sigma\sqrt{\gamma(Y(t);1,\nu) - \gamma(Y(t-h);1\nu)}z,
	\end{align*}
	where $z \sim N(0,1)$ and $\gamma(Y(t);1,\nu) \sim \Gamma(\frac{h}{\nu}, \nu)$. Therefore,
	\begin{align*}
		\gamma(Y(t);1,\nu) - \gamma(Y(t-h);1,\nu) &= \Gamma\bigg(\frac{Y(t)}{\nu}, \nu\bigg) - \Gamma\bigg(\frac{Y(t-h)}{\nu}, \nu\bigg)\\
		&= \Gamma\bigg(\frac{Y(t)-Y(t-h)}{\nu}, \nu\bigg).
	\end{align*}
	So we can write the differential as
	\[ \Delta Z(t) = \theta \Gamma\bigg(\frac{Y(t)-Y(t-h)}{\nu}, \nu\bigg) + \sigma\sqrt{\Gamma\bigg(\frac{Y(t)-Y(t-h)}{\nu}, \nu\bigg)}z. \]
	Milstein discretization of the CIR time change results in:
	\[ y_j = y_{j-1} + \kappa(\eta - y_{j-1})h + \lambda\sqrt{y_{j-1}h}z + \frac{\lambda^2}{4}h(z^2-1). \]
	This transforms the stochastic clock as $\int_{t_{j-1}}^{t_j}y(u)du$ and by the trapezoidal approximation we obtain
	\[ \int_{t_{j-1}}^{t_j}y(u)du = \frac{h}{2}(y_{j-1}-y_j). \]
	The change in the log asset price can be computed as follows. Given
	\begin{align*}
		\log S_t &= \log S_0 + (r-q)t + Z(t) - \log\mathbb{E}e^{Z(t)},\\
		\log S_{t-h} &= \log S_0 + (r-q)(t-h) + Z(t-h) - \log\mathbb{E}e^{Z(t-h)},
	\end{align*}
	subtracting yields the log stock price,	
	\[ \log S_t = \log S_{t-h} + (r-q)h + \Delta Z(t) + \Delta\omega_t. \]
	where $\omega = \log(\varphi_{VGSA}(-i \psi_{VG}(-i), (j-1)h, 1/\nu, \kappa, \eta, \lambda))-\log(\varphi_{VGSA}(-i\psi_{VG}(-i), jh, 1/\nu, \kappa, \eta, \lambda))$ is the martingale correction.
	Our algorithm for simulating the modified VG to include mean reverting time change is:
	\lstset{tabsize=3, language=MatLab, numbers=left, numbersep=4pt, frame=single, commentstyle=\color[rgb]{0,0.6,0}}
	\begin{lstlisting}[mathescape]
		S = 100; T = 5; r = .1; q = 0;
		%VG parameters
		sigma = 0.3; nu = 0.2; theta = -.14;
		%CIR parameters
		kappa = 0.01;    %Rate of mean reversion
		lambda = 0.02;    %Volatility of time change
		eta = 0.3;       %Long term rate of change
		
		K =linspace(10,150,50);
		numPaths = 1000;
		N = 100;    %number of time steps per path
		%VGSA Characteristic function
		phi_VGSA = @(u,t,y0,kappa,eta,lambda) (exp(kappa^2*eta*t/lambda^2)...
			/(cosh(sqrt(kappa^2-2*lambda^2*1i*u)*t/2)+kappa/sqrt(kappa^2...
			-2*lambda^2*1i*u)*sinh(sqrt(kappa^2-2*lambda^2*1i*u)*t/2))...
			^(2*kappa*eta/(lambda^2))*exp(2*1i*u/(kappa + sqrt(kappa^2-2...
			*lambda^2*1i*u)*coth(sqrt(kappa^2-2*lambda^2*1i*u)*t/2)))*y0);
		%Log VG characteristic function 
		psi_VG = @(u) (-1/nu*log(1-1i*u*theta*nu-sigma^2*nu*u^2/2));
		h = T/N;
		payoff = 0;
		for i = 1:numPaths
			X = 0;
			y = ones(1,N);
			lnS = log(S);
			for j = 2:N
				Z = normrnd(0,1);
				%Discretize the CIR time change
				y(j) = y(j-1) + kappa*(eta-y(j-1))*h+lambda...
					*sqrt(y(j-1)*h)*Z+lambda^2/4*h*(Z^2-1);
				tj = h*(y(j)+y(j-1))/2;
				G = gamrnd(tj/nu,nu); Z = normrnd(0,1);
				%CIR time changed VG process
				X = theta*G+sigma*sqrt(G)*Z;
				%Martingale correction
				omega = log(phi_VGSA(-1i*psi_VG(-1i), (j-1)*h, 1/nu, kappa,...
					eta, lambda))-log(phi_VGSA(-1i*psi_VG(-1i), j*h,...
					1/nu, kappa, eta, lambda));
				lnS = lnS + (r-q)*h+omega+X;
			end
			payoff = payoff + max(exp(lnS)-K,0);
		end
		payoff = exp(-r*T)*payoff/numPaths;
	\end{lstlisting}
	
	\subsection{Numerical Results of Monte-Carlo Simulation}
	Figure 3 shows the relative error between the individual Monte Carlo routines and the corresponding Fourier-cosine algorithms for VG, CGMY, and VGSA respectively. Each test involves 10,000 paths, uses a spot price of 100, strikes ranging from 50 to 150 and parameters $T = 1, r = 0.08, q = 0$. The Variance Gamma parameters are $\sigma = 0.41, \nu = 0.1, \theta = -0.1 $. CGMY uses $C = G = M = 10, Y = 1.5$. VGSA uses the same parameters for VG, with $\kappa = \eta = \lambda = 0.001$ which brings VGSA very close to a VG process, as we can see by the similar errors. 
	\begin{figure}[h!]
		\centering
	\includegraphics[scale=0.7]{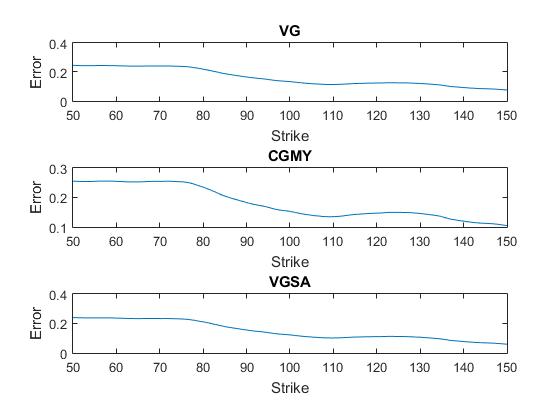}
	\caption{Error of Monte Carlo vs Fourier Cosine}
	\end{figure}
	\newpage
	\section{Least-Squares Calibration}
	In this section we analyze the first (of two) major methods of fitting option pricing models to market information.	Given a series of prices $\widehat{C}_i$ and strikes $K_i,\; i=1:n$, we search for the characteristic triplet of a risk-neutral exponential L\'{e}vy model $Q$ describing the model parameters under which the discounted asset price $e^{-rt}S(t)$ is a martingale. The measure $Q$ is chosen so that it minimizes the error with respect to the market prices. The calibration problem is the inverse of the option pricing problem. We construct L\'{e}vy process $Q$ so that the call option prices are given by their risk-neutral expectations
	\[ \widehat{C}_i \approx C^Q = e^{-rT}\mathbb{E}^Q\big[(S(T_i) - K_i)^+ \big]  \quad \text{for each $i$},\]
	and the discounted process is a martingale.
	
	The typical least squares problem is best explained by example. Given a series of market data (option prices and strikes) with the same maturity, risk-free rate, and dividend rate, we minimize the error between the observed prices and the ones predicted by our model $Q$. While there are many choices of error functional, throughout the subsequent examples we use the root mean square error, given by
	\[ RMSE(Q) = \frac{1}{\sqrt{N}} \sqrt{\sum_{i=1}^{N}(\widehat{C}(i) - C^Q(i))^2 }. \]
	In this section, we assume the underlier follows a VGSA process, and we construct an optimization problem to find the parameters which best fit the data.	Recall the VGSA process and corresponding asset price is defined as follows:
	\[ Z_{VGSA}(t) = X_{VG}(Y(t); \sigma, \nu, \theta) = b(\gamma(Y(t); 1, \nu), \sigma, \theta),\]
	where $\kappa, \lambda, \sigma, \nu, \theta, \eta$ are the six parameters defining the model. The characteristic function is given by
	\[ \mathbb{E}e^{iuZ_{VGSA}(t)} = \varphi(-i\Psi_{VG}(u), t, \nu^{-1}, \kappa, \eta, \lambda), \]
	where $\Psi_{VG}(u)$ is the log characteristic function of VG at unit time:
	\[ \Psi_{VG}(u) = -\frac{1}{\nu}\log(1-iu\nu\theta + \sigma^2\nu u^2/2). \]
	We may now define the asset pricing process at time $t$ as
	\[ S(t) = S(0)\frac{e^{(r-q)t+Z_{VGSA}(t)}}{\mathbb{E}e^{Z_{VGSA}(t)}}, \]
	where $T,r,q$ are obtained from market information, we must therefore determine the optimal $\sigma,\nu,\theta,\kappa,\eta, \lambda$. Any derivative-free nonlinear optimization method will suffice in determining a local minima. The images at the end of this section show the highly nonconvex error landscape for the VGSA naive least squares problem. 

	The typical least-squares optimization problem is ill-posed. We show a particularly interesting way to encourage convexity within the minimization functional and thus guarantee the existence of a solution. While the Black-Scholes model has to be replaced with models with finer structure such as those with jumps [27], the inverse problem is still ill-posed [65]. Various methods have been proposed which enforce some degree of stability, but restrict their domain to diffusion models [65,61,42]. Given the calibration problem's ill-posed nature, we must define extra criteria to ensure the models market price compatibility. Relative entropy as a criteria has solid foundations [24]. First we describe the notion of relative entropy.

	\subsection{Relative Entropy for L\'{e}vy Processes}

	As explained earlier, choosing an arbitrage free pricing model is equivalent to determining some measure $\mathbb{Q}$ satisfying the laws of the measure $\mathbb{P}$ under the constraint of being a martingale. There a many ways to determine the ``distance" between these two measures. We will focus on relative entropy, or Kullback Liebler divergence, which is defined by choosing the distance function $f(x) = x\ln x$ so that
	\[ \varepsilon(\mathbb{Q}, \mathbb{P}) = \mathbb{E}^\mathbb{Q}\bigg[ \frac{d\mathbb{Q}}{d\mathbb{P}}\bigg] = \mathbb{E}^\mathbb{P}\bigg[\frac{d\mathbb{Q}}{d\mathbb{P}}\ln\frac{d\mathbb{Q}}{d\mathbb{P}}\bigg]. \]
	Including the above requirement within our optimization problem results in the minimal entropy martingale measure. The intuition behind this measure $\mathbb{Q}$, or MEMM, is that it satisfies the martingale requirement while adding the least amount of information to the prior model $\mathbb{P}$. In the case of exponential L\'{e}vy models, the MEMM does not always exist, but we can determine analytically the criterion for its tractability. The MEMM can also be related to the Escher transform [20]. Further reading on the concept of MEMM can be found in [33, 56, 23, 30]. 
	
	The relative entropy between two L\'{e}vy processes can be thought of as the disparity of information contained within them and is thus an effective measure of their distance [17]. Relative Entropy for L\'{e}vy Processes is defined as follows [20]. First, let $\{X_t\}_{t\ge0}$ be a real-valued L\'{e}vy process defined on spaces $(\Omega, \mathcal{F}, \mathbb{P}), (\Omega, \mathcal{F}, \mathbb{Q})$ with respective characteristic triplets $(A_P, \nu_P, \gamma_P),(A_Q, \nu_Q, \gamma_Q) $. Suppose that $\mathbb{Q} \ll \mathbb{P}$ (PIIS Thm [43]) which implies that $A_Q = A_P, \nu_Q\ll \nu_P$ and define $A\equiv A_Q=A_P$, then for every time horizon $T\le T_\infty$, the relative entropy of $\mathbb{Q}|_{\mathcal{F}_T}$ with respect to $\mathbb{P}|_{\mathcal{F}_T}$ is computed as
	\begin{align*}
		\varepsilon(\mathbb{Q},\mathbb{P}) &= \varepsilon(\mathbb{Q}|_{\mathcal{F}_T},\mathbb{P}|_{\mathcal{F}_T})\\
		&= \frac{T}{2A}\bigg[ \gamma_Q - \gamma_P - \int_{-1}^{1}x(\nu_Q - \nu_P)dx \bigg]^2\mathbb{I}_{A\ne 0} + T\int_{-\infty}^{\infty}\bigg( \frac{d\nu_Q}{d\nu_P}\log\frac{d\nu_Q}{d\nu_P} + 1 - \frac{d\nu_Q}{d\nu_P} \bigg)\nu_Pdx.\tag{1}
	\end{align*}
	The relative entropy criterion satisfies the following properties: [59]
	\begin{enumerate}
		\item Convexity: with two probability measures $\mathbb{Q}_1, \mathbb{Q}_2$ both equivalent to $\mathbb{P}$ then
		\begin{align*}
		\varepsilon(\alpha \mathbb{Q}_1 + (1-\alpha)\mathbb{Q}_2, \mathbb{P}) &= \mathbb{E}^\mathbb{P}\bigg[ f\bigg( \frac{d(\alpha \mathbb{Q}_1 + (1-\alpha)\mathbb{Q}_2)}{d\mathbb{P}}\bigg)\bigg]\\
		&= \mathbb{E}^\mathbb{P}\bigg[ f\bigg(\alpha \frac{d\mathbb{Q}_1}{d\mathbb{P}} + (1-\alpha)\frac{d\mathbb{Q}_2}{d\mathbb{P}}\bigg)\bigg]\\
		&\le \alpha \varepsilon(\mathbb{Q}_1,\mathbb{P}) + (1-\alpha) \varepsilon(\mathbb{Q}_2,\mathbb{P}).
		\end{align*}
		\item Nonnegativity: $\varepsilon(\mathbb{P}, \mathbb{Q})\ge 0$.
		\item $\varepsilon(\mathbb{Q}, \mathbb{P}) = 0 \iff \frac{d\mathbb{Q}}{d\mathbb{P}} = 1$ a.s..
	\end{enumerate}

	\subsection{Modified Least-Squares Functional}

	The minimal relative entropy approach was extended to stochastic processes through the weighted Monte Carlo method by Avellaneda [6] and while Goll and R\"{u}schendorf[34] describe calibration of minimal relative entropy, they do not propose an algorithm. We follow the method used by Cont and Tankov [20] which overcomes the shortcomings of the above approaches by defining the calibration problem to include jump processes and use relative entropy as a regularization criterion instead of a selection criterion. We identify the L\'{e}vy measure $\nu$ and volatility $\sigma$ from observation of liquid call prices. Given adequate information, their determination is as follows:
	\begin{enumerate}[noitemsep]
		\item Determine the risk-neutral density using the Breeden-Litzenberger formula [12]
				\[ p_X(t) = e^{r(T-t)}\frac{\partial^2}{\partial K^2}C(t,S(t)). \]
		\item Compute characteristic function by Fourier transform of the density $p_X$.
		\item $\sigma, \nu$ can be determined from the characteristic function through Fourier inversion
	\end{enumerate}
	We first construct the norm
	\[ \|C\|^2_w = \int_{\mathfrak{C}} C(T,K)^2w(dT\times dK),  \]
	where the probability weighting measure $w=\sum_{i=1}^{N}w_i\delta_{(T_i, K_i)}(dT\times dK)$ corresponds to the weight of each individual constraint $1:N$ and $w$ is defined on the grid of strikes and maturities $\mathfrak{C} = [0, T_\infty]\times[0,\infty)$. Thus the quadratic pricing error for $Q$ is
	\[ \|\widehat{C} - C^Q \|^2_w = \sum_{i=1}^{N}w_i(\widehat{C}(T_i, K_i) - C^Q(T_i, K_i))^2. \]
	The non-linear least-squares calibration problem can be formulated as
	\[ \|\widehat{C} - C^{Q^*}\|^2_w = \min_{Q\in \mathcal{M}\cap\mathcal{L}} \|\widehat{C} - C^Q \|^2_w   \]
	where the model $\mathbb{Q}^*$ is the least squares solution that minimizes the squared pricing errors between the model and the market. $\mathbb{Q}^*$ lies in the set of all L\'{e}vy martingale measures $\mathcal{M}\cap\mathcal{L}$. This problem, however, is ill-posed since we cannot guarantee the existence of a solution, the least-squares functional is non-convex which prevents any gradient-based algorithm from finding the optimal solution, and it is highly sensitive to initial conditions.

	Since models involving jumps do not always admit a unique measure, integrating prior views into the calibration procedure allows us to determine some martingale measure that expresses the observed option prices while maintaining equivalence to the historical measure or the preconceived notions of the investment manager. In this section we calibrate the problem with respect to a prior model which is the product of historical prices. Since the calibration procedure is likely to be executed daily, a good choice of prior is the optimal measure of the last time the calibration procedure was executed.

	\subsection{Regularization}

	We enforce uniqueness, existence, and stability into our calibration problem by introducing additional information through a prior model $\mathbb{P}$. We aim to find a solution to the least-squares calibration problem which minimizes the relative entropy with respect to our prior model $\mathbb{P}$. Given two probability measures $\mathbb{P}$ and $\mathbb{Q}$ with respective Levy triplets $(\sigma^2, \gamma_P, \nu_P)$ and $(\sigma^2, \gamma_Q, \nu_Q)$, we define our regularization term as the relative entropy of $\mathbb{Q}$ with respect to $\mathbb{P}$, 
	\[ \varepsilon(\mathbb{Q}, \mathbb{P})= \frac{T}{2\sigma^2} \bigg[ \int_{-\infty}^{\infty} (e^x - 1)(\nu_Q - \nu_P)dx\bigg]^2 + T\int_{-\infty}^{\infty} \bigg[ \frac{d\nu_Q}{d\nu_P}\ln \frac{d\nu_Q}{d\nu_P} + 1- \frac{d\nu_Q}{d\nu_P} \bigg] \nu_Pdx   \]
	We follow the approach outlined in [28] for regularization of ill-posed problems and add the minimal relative entropy $\varepsilon(\mathbb{Q}, \mathbb{P})$ as a penalization term so that our least squares problem takes the following form: 
	\[   J_\alpha(\mathbb{Q}^*) = \min_{Q\in\mathcal{M}\cap\mathcal{L}} \|\widehat{C} - C^Q \|^2_w  + \alpha\varepsilon(\mathbb{Q}, \mathbb{P}).  \]
	If the regularization parameter $\alpha$ is large, the functional becomes convex and approaches the prior which enforces stability. If $\alpha$ is small, the functional is close to the least-squares problem which enforces precision. An appropriate choice of regularization parameter is necessary to the accuracy of the solution, as we will see later. The previous equation, through slight abuse of language, is the regularized minimal entropy martingale calibration problem (RMEMC).
	
	\subsection{Numerical Implementation of RMEMC}

	As we have shown, the initial calibration problem can be reformulated as an optimization problem determining the L\'{e}vy measure $\mathbb{Q}^*$ representing the minimum value of the regularization problem
	\[ J_\alpha(\mathbb{Q}^*) = \min_{Q\in\mathcal{M}\cap\mathcal{L}} \sum_{i=1}^{N}w_i(\widehat{C}(T_i, K_i) - C^Q(T_i, K_i))^2 + \alpha\varepsilon(Q, \mathbb{P}),  \]
	which can be solved once we obtain the
	\begin{enumerate}[noitemsep]
		\item constraint weights $w_i$,
		\item prior measure $\mathbb{Q}$, and
		\item regularization parameter $\alpha$.
	\end{enumerate}
	\nl
	\subtitle{Minimization Constraint Weights}	
	\nl 
	The weights within the optimization problem identify our ``confidence" in individual market data points. Options with higher trading volume (higher liquidity) will thus have a better estimation of their respective prices. As shown in [18], a reasonable solution is to minimize the squared differences between prices with respect to option vega, which is defined as the derivative of the Black Scholes option price with respect to volatility:
	\[ \text{vega} = |Ke^{-rT}N(d_-)\sqrt{T}|. \]
	After letting each weight be the square reciprocal of vega, our calibration problem becomes
	\[ J_\alpha(\mathbb{Q}^*) = \min_{Q\in\mathcal{M}\cap\mathcal{L}} \sum_{i=1}^{N}\frac{(\widehat{C}(T_i, K_i) - C^Q(T_i, K_i))^2}{(K_ie^{-rT_i}N(d_-)\sqrt{T_i})^2} + \alpha\varepsilon(Q, P).  \]
	\nl
	\subtitle{Determination of Prior Measure}
	\nl
	We aim to construct the prior measure automatically from option price data. As we will see, while the initial prior model determines the speed of convergence, the solution will still converge to optimality regardless of our initial choice of prior. For this reason, we let the prior model $\mathbb{P}$ for the calibration problem of determining $\mathbb{Q}_n$ to be equal to the optimal measure generated previously, ie. $\mathbb{P} = \mathbb{Q}_{n-1}$. This gives us the recursive relation
	\[ J_\alpha(\mathbb{Q}^*_n) = \min_{Q\in\mathcal{M}\cap\mathcal{L}} \sum_{i=1}^{N}\frac{(\widehat{C}(T_i, K_i) - C^Q(T_i, K_i))^2}{(K_ie^{-rT_i}N(d_-)\sqrt{T_i})^2} + \alpha\varepsilon(\mathbb{Q}_n, \mathbb{Q}_{n-1}).  \]
	This is solved using a derivative free gradient based procedure such as Nelder-Mead [9].\\
	\nl
	\subtitle{Regularization Parameter}
	\nl
	The authors of [20] suggest the optimal $\alpha$ be found using the Morozov discrepancy principle [28]. Briefly described, any parameter choice rule $\delta \mapsto \alpha (\delta)$, as the noise level $\delta \rightarrow 0$, must satisfy:
	\begin{enumerate}[noitemsep]
		\item $\alpha(\delta)\rightarrow0$,
		\item $\frac{\delta^2}{\alpha(\delta)} \rightarrow0$ given exactly attainable constraints,
		\item or, $\frac{\delta}{\alpha(\delta)} \rightarrow0$ if the model does not exactly replicate the data, as is often the case.
	\end{enumerate}
	The discrepancy principle was developed by Morozov to use on least squares regularization in Banach spaces [57,58] and it has been shown that such a method of choosing the regularization parameter yields the most numerically favorable results [68]. Using the assumptions that:
	\begin{enumerate}
		\item The prior model $\mathbb{Q}^*$ is arbitrage free satisfying minimal entropy martingale measure criteria,
		\item There exists a solution $\mathbb{Q}^+$ to the non regularized problem (MEMC) with finite entropy satisfying fully attainable constraints, ie $\|C^{Q^+}-\widehat{C}\|_w=0$, and
		\item There exists maximum noise level $\delta_0$ such that the max error satisfies $\epsilon^{\text{max}}:=\inf_{\delta\le\delta_0}\|C^{Q^*}-\widehat{C}^\delta\|^2_w>\delta_0^2$
	\end{enumerate}
	Where $C^\delta_0$ represents perturbed price data within some noise level $\delta$ of $\widehat{C}$. Because this noisy data will not yield a very good solution from MEMC, we need to regularize the problem by defining a family of regularization operators $\{R_\alpha\}_{\alpha\ge0}$ where $\alpha$ determines the intensity of the regularization [68]. If $\alpha$ is chosen appropriately the regularized, noisy problem converges to the MEMC problem admitting an exact solution, ie. RMEMC$_\alpha(\widehat{C}^\delta)\rightarrow $MEMC$(\widehat{C})$ as $\delta\rightarrow 0$.	Denote $Q^\delta_\alpha$ to be the solution to the noisy problem RMEMC$_\alpha(C_M^\delta)$ with regularization parameter $\alpha$. The RMEMC a priori multivalued discrepancy function is thus
	\[ \epsilon_\delta(\alpha) := \|C^{Q^\delta_\alpha}-\widehat{C}^\delta\|^2_w. \]
	We place the following constraints on the discrepancy principle. Given two constants $c_1$ and $c_2$ satisfying
	\[ 1<c_1\le c_2 < \frac{\epsilon^{\max}}{\delta^2_0}, \]
	the discrepancy principle can be stated as, for a given noise level $\delta$, choose parameter $\alpha>0$ satisfying
	\[ \delta^2<c_1\delta^2\le \epsilon_\delta(\alpha)\le c_2\delta^2. \]
	We aim to find a solution $\mathbb{Q}$ of MEMC$(\widehat{C})$ with noise level of order $\delta$. We try to solve $\|C^{Q^\delta_\alpha}-\widehat{C}^\delta\|^2_w\le \delta^2$. By sacrificing some precision we gain stability so we pick some constant $c\approx1^+$ (e.g. $c=1.1$) and search for the solution $\mathbb{Q}^\delta_\alpha$ in the level set defined by
	\[ \|C^{Q^\delta_\alpha}-\widehat{C}^\delta\|^2_w\le c\cdot \delta^2, \]
	where the highest stability is achieved when $\epsilon_\delta(\alpha) := \|C^{Q^\delta_\alpha}-\widehat{C}^\delta\|^2_w = c\cdot \delta^2 $
	The noise level $\delta$ can be computed directly if the bid and ask prices are known, ie. given:
	\[ \widehat{C}^\delta(T_i, K_i) = \frac{\widehat{C}^{\text{bid}}(T_i, K_i) + \widehat{C}^{\text{ask}}(T_i, K_i)}{2},\qquad \forall i=1:N, \]
	the noise level is thus:
	\[ \delta := \frac{\|\widehat{C}^{\text{bid}} + \widehat{C}^{\text{ask}}\|_2}{2}. \]
	While the theoretical foundations of using the discrepancy principle are solid, in practice it is unrealistic to determine the bid-ask spread of the options since liquidity may not offer reliable estimates. Instead, during the construction of our RMEMC algorithm, we opt for a similar approach of picking the regularization parameter based upon the size of vega, which as explained earlier is another method of estimating volatility of the markets. In fact, defining the regularization term as
	\[ \alpha = \mathcal{A}\cdot |Ke^{-rT}N(d_-)\sqrt{T}|, \quad \text{for some }\mathcal{A} > 0\]
	has the advantage of speeding up convergence when the volatility is high and restricting the incorporation of new market information (noise) when volatility is low. The term $\mathcal{A}$ represents the investors preconceived notions about the direction or fluidity of the market, and could just be set to 1. For instance, high volatility suggests that the market is responding to new information, and thus new price movements will likely have greater impact on the optimal pricing measure than historical price movements. When volatility is low, it is difficult to distinguish new information from noise, and so the regularization term acts as a dampening parameter which restricts change in the optimal measure.\\
	\nl
	\subtitle{Calibration Functional}
	\nl
	We first discretize the calibration problem by taking L\'{e}vy process $P$ with finite measure [68]:
	\[ \nu_P = \sum_{k=0}^{M-1}p_k\delta_{x_k}(dx), \]
	which satisfies
	\[ \nu_Q = \sum_{k=0}^{M-1}q_k\delta_{x_k}(dx) \ll \nu_P.\]
	Substituting the above expressions into the RMEMC problem yields, with for simplicity, $\mathbb{Q} = \mathbb{Q}_n$ and $\mathbb{P} = \mathbb{Q}_{n-1}$:
	\begin{align*}
		J_\alpha(\mathbb{Q}^*) :&= \|\widehat{C} - C^Q\|^2_w + \alpha \varepsilon(\mathbb{Q}|\mathbb{P})\\
		&= \|\widehat{C} - C^Q\|^2_w + \alpha \Bigg(\frac{1}{2A}\bigg[ \gamma_Q - \gamma_P - \int_{-1}^{1}x(\nu_Q - \nu_P)dx \bigg]^2\mathbb{I}_{A\ne 0} + \int_{-\infty}^{\infty}\bigg[ \frac{d\nu_Q}{d\nu_P}\log\frac{d\nu_Q}{d\nu_P} + 1 - \frac{d\nu_Q}{d\nu_P} \bigg]\nu_Pdx \Bigg)\\
		&= \sum_{i=1}^{N}\frac{(\widehat{C}(T_i, K_i) - C^Q(T_i, K_i))^2}{(K_ie^{-rT_i}N(d_-)\sqrt{T_i})^2} + \frac{\alpha}{2A}\bigg(\frac{A}{2}+b_P + \sum_{j=0}^{M-1}(e^{x_j}-1)q_j\bigg)^2 + \alpha\sum_{j=0}^{M-1}\bigg[q_j\log\bigg(\frac{q_j}{p_j}\bigg) + 1-q_j\bigg]
	\end{align*}
	Where $b_P=\gamma_P-\int_{-1}^{1}x\nu_P(dx)$ is the drift of the prior $\mathbb{P}$.

	\begin{figure}[!h]
		\begin{subfigure}[h]{0.5\textwidth}
			\includegraphics[scale=0.46]{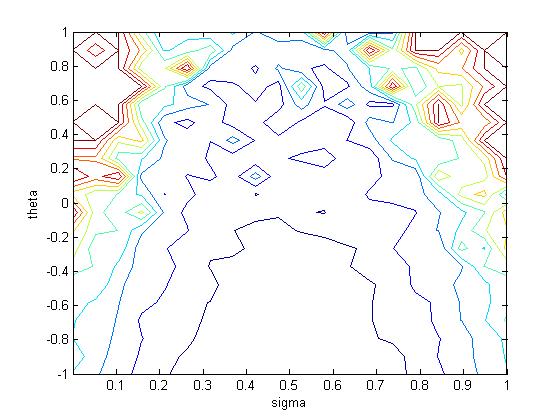}	
			\caption{VGSA Error Contour}
		\end{subfigure}
		\begin{subfigure}[h]{0.5\textwidth}		
			\includegraphics[scale=0.46]{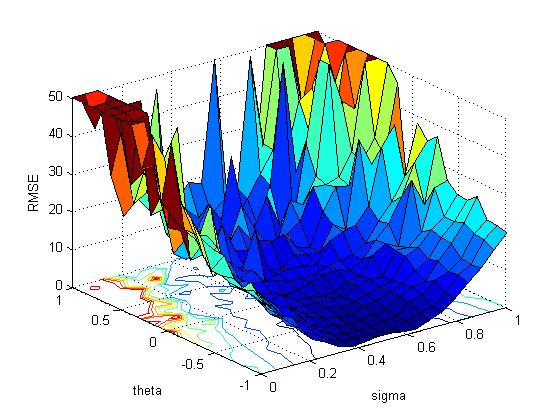}
			\caption{VGSA Error Surface}
		\end{subfigure}
		\caption{Naive Least Squares Error Surface}
	\end{figure}
	\begin{figure}[!h]
		\begin{subfigure}[h]{0.5\textwidth}
			\includegraphics[scale=0.46]{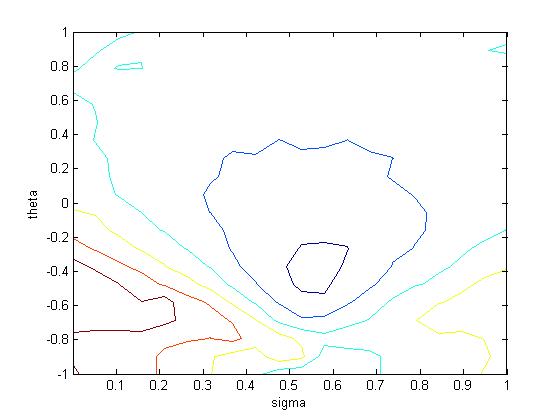}	
			\caption{RMEMC Contour}
		\end{subfigure}
		\begin{subfigure}[h]{0.5\textwidth}
			\includegraphics[scale=0.46]{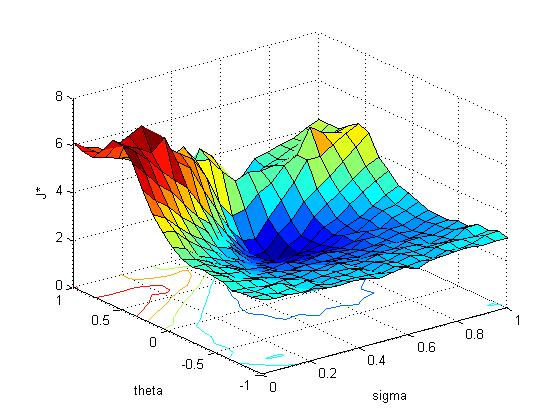}
			\caption{RMEMC Surface}
		\end{subfigure}
		\caption{Regularized Minimal Entropy Functional}
	\end{figure}
	The images of Figures 3 and 4 show exactly how nonconvex the naive least squares error surface is. The act of regularizing the error metric with respect to the minimal entropy martingale measure enforces convexity, but does not guarantee convexity. That is what the regularization parameter $\alpha$ is for. For $\alpha=0$ we obtain the naive least squares problem. For some $\alpha>0$ we obtain a function that is strictly convex, but our new parameter estimates will become closer to that of the prior measure. Careful choice of $\alpha$ must be made to ensure existence and stability while also reducing the error of the regularized pricing measure with respect to that of the prior. Obtaining a new measure with error greater than that of the prior is clearly undesirable, which will be avoided by limiting the size of $\alpha$. Ensuring that the pricing error decreases with use of RMEMC, we may have to settle with partial regions of nonconvexity in the RMEMC surface, but this can be overcome. A statistical heuristic for determining multiple starting points lends the minimization problem especially well to parallelization and, combined with the increased convergence rate of RMEMC, will significantly increase the capacity of finding a global optimum.
	\subsection{RMEMC Algorithm}
	\lstset{tabsize=3, language=MatLab, numbers=left, numbersep=4pt, frame=single, commentstyle=\color[rgb]{0,0.6,0}}
	\begin{lstlisting}[mathescape]
		data = xlsread('SPY Options', 'Prices', 'A2:B69');
		K = data(:,1);
		MPrice = data(:,2);
		n = length(K);
		
		S0 = 90.692;
		T = 0.194387;
		r = 0.0179;
		q = 0;
		
		%Determine prior
		start = [0 -1 -1];
		options = optimset('MaxFunEvals', 1e10, 'MaxIter', 1e3, 'TolX', 1e-5);
		[prior, ~] = fminsearch(@(b) cal_RMSE(b, S0, K, MPrice, r, q, T),...
			start, options);
		
		theta = prior(1);
		sigma = prior(2);
		nu = prior(3);
		%Use volatility derivative as weights
		vega = @(S, T, K, sigma) (S*normpdf(1/sqrt(T)*(log(S./K)...
			+1/2*sigma^2*T))*sqrt(T));
		w = 1./vega(S0, T, K, sigma);
		alpha = 0.03;
		
		charvec = [-50:50];
		b = zeros(1,3);
		%RMEMC functional
		J = @(b, S0, K, MPrice, r, q, T) (cal_RMSE(b,S0,K,MPrice,r,q,T, w))+...
			alpha*RelativeEntropy(cf_VG(charvec,T,r,b(1),b(2),b(3)),...
			cf_VG(charvec,T,r,sigma,nu,theta));
		
		[reg, ~] = fminsearch(@(b) J(b,S0,K,MPrice,r,q,T), start, options);
		C = zeros(1,n);
		D = zeros(1,n);
		for i=1:n
			C(i) = cos_VG(S0, K(i), r-q, T, theta, sigma, nu);
			D(i) = cos_VG(S0, K(i), r-q, T, reg(1),reg(2),reg(3));
		end
		
		plot(K, C, 'd', K, MPrice, '.', K, D, '*');
	\end{lstlisting}
		%
		%
		%***********************************************************************************************
		%***********************************************************************************************
		%
		%

	\section{Parameter Estimation}

	In the previous section we obtained deterministically the model which minimizes the error between estimated and actual market prices. The calibration procedure utilizes cross-sectional instruments while disregarding time series information. Parameter estimation develops the optimal pricing model in a supplemental fashion by incorporating historical asset movements. Realistic applications often require sample path generation of an unknown distribution inferred by some maximum likelihood or sequential Monte Carlo approach. Maximum likelihood is consistent and guaranteed to converge to the true distribution over time [66]. In cases of singular noise such as GARCH [29, 10, 21] the likelihood function is available in integrated form. For partially observed processes, this is not the case, and a filtering technique is required. Partially observed processes do not have an explicit integrated density function and so we are forced to utilize the conditional density by calculating the hidden state (variable) on that day to best describe the observation.
	\subsection{Filtering}
	We first give an overview of the concept of filtering. It is an iterative process enabling us to determine model parameters using historical time-series data. The idea is to construct a \textit{transition} equation connecting consecutive hidden states and a \textit{measurement} equation connecting the hidden state to the observable data. We first estimate the hidden state at time $t$ \textit{a priori} using all prior information up to time $t-1$. We then construct a conditional \textit{a posteriori} estimate using the prior measurement and the current time $t$ observation. The main concept behind filtering is summarized by the following two steps which are executed recursively.
	\begin{enumerate}
		\item \textbf{Prediction (time) update:} Given observations up to time $t_{k-1}$, we apply the Chapman-Kolmogorov equation to determine the best prediction for $x_k$ at time $t_k$ described by the prior density
		\[ p(x_k|z_{1:k-1}) = \int p(x_k|x_{k-1})p(x_{k-1}|z_{1:k-1})dx_{k-1}. \]
		\item \textbf{Measurement update:} Given observation $z_k$ we apply Bayes rule to determine the probability of $x_k$ to determine the posterior density
		\[ p(x_k|z_{1:k}) = \frac{p(z_k|x_k)p(x_k|z_{1:k-1})}{p(z_k|z_{1:k-1})}. \]
	\end{enumerate}
	A simple yet very effective example of filtering is the Kalman filter. First developed for use in control engineering and signal processing, they have been vital toward the implementation of tracking systems, navigation, guidance systems, and most notably the trajectory estimation system of the Apollo program. The traditional Kalman filter is applicable only to linear systems. We will use the extended Kalman Filter (EKF) which is based upon a first order linearization of the transition and measurement equations which admits application to nonlinear systems. 
	\subsection{Extended Kalman Filter}
	The extended Kalman filter is used in the special case where the proposal density $p(x_k|z_{k-1})$ and observation density $p(x_k|z_k)$ are nonlinear and Gaussian [44]. They are modeled by a Markov chain built on operators perturbed by Gaussian noise. We assume a dynamic process $x_k$ follows the nonlinear transition equation
	\begin{align*}
		x_k &= f(x_{k-1},w_k),\\
		w_k &\sim N(0,Q_k).
	\end{align*}
	Suppose a measurement $z_k$ follows the observation equation
	\begin{align*}
		z_k &= h(x_k,u_k),\\
		u_k &\sim N(0,R_k).
	\end{align*}
	The terms $w_k,u_k$ are two mutually independent sequences of uncorrelated normal random variables representing the process noise and observation noise with covariance matrices $Q_k, R_k$ respectively. Define the a priori estimate of the system given all except the current observation as
	\[ \widehat{x}_{k|k-1} = \mathbb{E}[x_k|x_{k-1}]. \]
	Define the a posteriori estimate given all current information as
	\[ \widehat{x}_{k|k} = \mathbb{E}[x_k|x_k]. \]
	Then, we can define the a priori and a posteriori error covariance matrices as
	\begin{align*}
		P_{k|k-1} &= \text{cov}(x_k-\widehat{x}_{k|k-1}),\\
		P_{k|k} &= \text{cov}(x_k-\widehat{x}_{k|k}).
	\end{align*}
	It is here where the EKF and traditional KF systems differ. We construct a linearization of the transition and measurement equations; the EKF reduces to the KF when the equations actually are linear. The Jacobian matrices of the transition equation with respect to the system process, $A_k$, and system noise, $W_k$, are defined as
	\[ A_{ij} = \frac{\partial f_i}{\partial x_j}(\widehat{x}_{k|k-1},0) \qquad\text{and}\qquad W_{ij} = \frac{\partial f_i}{\partial w_j}(\widehat{x}_{k|k-1},0). \]
	Similarly, the Jacobians of the observation equation with respect to system process $H_k$ and measurement noise $U_k$ are
	\[ H_{ij} = \frac{\partial h_i}{\partial x_j}(\widehat{x}_{k|k},0) \qquad\text{and}\qquad U_{ij} = \frac{\partial h_i}{\partial w_j}(\widehat{x}_{k|k},0). \]
	This gives us the following time update equations for the a priori state estimate and error covariance
	\begin{align*}
		\widehat{x}_{k|k-1} &= f(\widehat{x}_{k-1|k-1},0),\\
		P_{k|k-1} &= A_kP_{k-1}A_k^T+W_kQ_{k-1}W_k^T,
	\end{align*}
	and measurement update equations for the respective a posteriori variables
	\begin{align*}
	\widehat{x}_{k|k} &= \widehat{x}_{k|k-1} + K_k(z_k - h(\widehat{x}_{k-1},0)),\\
	P_{k|k} &= (I-K_kH_k)P_{k|k-1}.
	\end{align*}
	The term $K_k$ represents the optimal Kalman gain matrix found by minimizing the mean square error, or the trace of $P_{k|k}$, over all linear estimators
	\[ K_k = P_{k|k-1}H_k^T(H_kP_{k|k-1}H_k^T + U_kR_kU_k^T)^{-1}. \]
	The Kalman gain corresponds to the mean of the conditional distribution of $x_k$ given observation $z_k$.
	
	Because of the first order linearization required by the EKF algorithm, it is known to fail in substantially nonlinear systems or if the state equations are highly non-Gaussian. An early example of an attempt to correct this shortcoming was to approximate the posterior by expansion in a prespecified function basis known as the Gaussian sum filter [2,47]. Several algorithms have been described which use a deterministic set of points to represent the posterior distribution such as the unscented Kalman Filter (UKF) [45,69] and Gaussian quadrature Kalman Filter (QKF) [40]. These methods have the advantage of not having to compute the Jacobian matrix which is often the most computationally expensive step within the EKF paradigm.
	
	These methods all suffer if the posterior density can be represented by a Gaussian distribution, which is often not the case. Many attempts at successfully overcoming these limitations require the use of Monte Carlo methods to represent the posterior by a collection of random points. The uses of Monte Carlo for filtering of nonlinear processes can be traced to (Gordon et al 1993) [40,36] and were based off sequential importance sampling (SIS). This technique required simulating samples under some proposal distribution and then approximating the target distributions by applying appropriately defined importance weights. The sequential nature of SIS results from, in a nonlinear filtering context, defining the appropriate sequence of state distributions such that regenerating the sample population upon each new observation becomes unnecessary. SIS is a method described more generally as particle filtering.
	\subsection{Particle Filtering}
	Also known as sequential Monte Carlo (SMC), particle filtering is a recent alternative [5] for parameter estimation of nonlinear processes by discretizing the continuous density function $p(x_t|y_t)$. For $N$ sample points we obtain a sequence $\{x_t^{(i)}, w_t^{(i)} \}_{i=1}^N$ where $w_t^{(i)}$ is the weight associated to each particle $x_t^{(i)}$ at time $t$. The expectation with respect to the filter is
	\[ \mathbb{E}f(x_t) = \int f(x_t)p(x_t|y_t)dx_t \approx \sum_{i=1}^{N}w_t^{(i)}f(x_t^{(i)}). \]
	The first step in filtering is determining the initial value. Our approach is to use the extended Kalman filter to converge quickly to the prior state value. The optimal proposal distribution is given by the target distribution
	\[\pi(x_k|x_{0:k-1}, y_{0:k}) = p(x_k|x_{k-1}, y_{k}). \]
	In practice, the transition prior is often used as the importance function which yields
	\[\pi(x_k|x_{0:k-1}, y_{0:k}) = p(x_k|x_{k-1}). \]
	We associate the signal noise with $R_v$ and the probability function becomes
	\[ p(y_{j,k}|x_k) = \frac{1}{\sqrt{2\pi R_{v,j,j}}}e^{-\frac{(y_{j,k}-F_j(x_k))^2}{2R_{v,j,j}}}. \]
	The Monte Carlo approximation of the likelihood at step $k$ is:
	\[ l_k = \sum_{i=1}^{N}\frac{p(y_k|x_{i,k})p(x_{i,k}|x_{i,k-1})}{\pi (x_{i,k}|x_{i,k-1},y_k)} = \sum_{i=1}^{N}w_k^{(i)}.\]
	Then, to estimate the parameters we minimize the negative log likelihood
	\[-\sum_{i=1}^{N}\log l_k.\]
	The SIS method suffers from a major drawback not properly identified until [34]. The importance weights degenerate over time, known as \textit{weight degeneracy}. The importance weights of most of the samples of the target distribution decrease over time to the point where they do not significantly contribute to the process. The bootstrap filter proposed by [34] solves this issue by regenerating the set of samples with importance weights above a prespecified threshold and disregarding those below it. This was the first successful attempt of applying SMC to nonlinear filtering [13]. Since then, various alternatives have been proposed such as the \textbf{con}ditional \textbf{dens}ity propag\textbf{ation} (condensation) filter [9], Monte Carlo filter [46] and sequential imputations [48]. The bootstrap filter, or sequential importance resampling (SIR) samples $N$ draws from the current set of particles using the normalized weights as individual selection probabilities. Trajectories with small weights are eliminated, and those with larger weights are replicated. The standard particle filtering algorithm generally refers to the SIR method.
	
	\subsection{Sequential Importance Resampling}
	As mentioned earlier, to prevent weight degeneracy causing algorithmic divergence [5] we regenerate particles with higher weight and eliminate those with lower weight. The SIR particle filtering algorithm is described as follows.
	\begin{enumerate}
		\item Simulate the state from the prior by drawing $N$ samples according to the model
		\[ x_k^{(i)}= f(x_{k-1}^{(i)}, u_{k-1}^{(i)}),\quad i=1:N. \]
		\item Associate the weights for each point by updating the importance function
		\[ w^{(i)}_k = w_{k-1}^{(i)} \frac{p(z_k|x_k^{(i)})p(x_k^{(i)}|x_{k-1}^{(i)})}{\pi(x_k^{(i)}|x_{k-1}^{(i)}, z_k)}. \]
		Letting the proposal distribution $\pi(x_k^{(i)}|x_{k-1}^{(i)}, z_k)$ equal the transition probability $p(x_k^{(i)}|x_{k-1}^{(i)}),$ the expression simplifies to
		\[ w_k^{(i)} = w_{k-1}^{(i)}p(z_k|x_k^{(i)}). \]
		\item Normalize the weights
		\[ \widehat{w}_k(x_k^{(i)}) = \frac{w_k^{(i)}}{\sum_{i=1}^{N}w_k^{(i)}}. \]
		\item Resample the weights: compare the CDF of the normalized weights to a uniform CDF, if
		\[ \frac{1}{N}(U(0,1)+j-1)\ge \sum_{l=1}^{i}\widehat{w}_k(x_k^{(l)}), \]
		then increment and skip $i$, else set the weight $\widehat{w}_k(x_k^{(i)}) =\frac{1}{N}$.		
	\end{enumerate}
	Therefore, the best estimate of $x_k$ is the conditional expectation
	\[ \mathbb{E}[x_k|z_{1:k}] \approx \sum_{i=1}^{N}\widehat{w}_k(x_k^{(i)})x_k^{(i)}. \]

	\subsection{Integrated Densities}
	
	For a fully observed process, such as Variance Gamma, its probability density function is available in integrated form. Recall the VG process is defined as
	\[ X(t;\sigma,\nu,\theta) = \theta\gamma(t;1,\nu) + \sigma W(\gamma(t;1,\nu)), \]
	with log asset price
	\[ \ln S_t = \ln S_0 + (r-q+\omega)t + X(t;\sigma,\nu,\theta), \]
	where $\omega = \frac{1}{\nu}\ln (1-\theta\nu - \sigma^2\nu/2)$ is the VG martingale correction. Given the following definitions:
	\begin{align*}
	x_h &= z_k - (r-q)h - \frac{h}{\nu}\ln (1-\theta\nu - \sigma^2\nu/2),\\
	z_k &= \ln\bigg(\frac{S_k}{S_{k-1}}\bigg),\\
	h &= t_k - t_{k-1} 
	\end{align*}
	we obtain the integrated VG density function
	\[ p(z_k|z_{1:k-1}) = \frac{2e^{\theta x_h/\sigma^2}}{\nu^{\frac{h}{\nu}}\sqrt{2\pi}\sigma\Gamma(\frac{h}{\nu})}\bigg(\frac{x^2_h}{2\sigma^2/\nu+\theta^2}\bigg)^{\frac{h}{2\nu}-\frac{1}{4}}K_{\frac{h}{\nu}-\frac{1}{2}}\bigg(\frac{1}{\sigma^2}\sqrt{x^2_h(2\sigma^2/\nu+\theta^2)}\bigg ) \]
	where $K_n$ is the modified Bessel function of the second kind [51].
			
	A partially observed process, such as VGSA, has a density function that is not available in closed form, therefore we condition on a hidden parameter to determine its conditional density function which is available in integrated form.\\
	Recall the log asset price following a VGSA process is defined as
	\begin{align*}
	d\ln S_t &= (r-q+\omega)dt + X(h(dt);\sigma,\nu,\theta),\\
	X(h(dt);\sigma,\nu,\theta) &= B(\gamma(h(dt),1,\nu);\theta,\sigma)
	\end{align*}
	where the gamma cumulative distribution function is
	\[ F_\nu(h,x) = \frac{1}{\Gamma(\frac{h}{\nu})\nu^\frac{h}{\nu}}\int_{0}^{x}e^{-\frac{t}{\nu}}t^{\frac{h}{\nu}-1}dt. \]
	The gamma time change is modeled by an integrated CIR process so we set
	\begin{align*}
	h(dt) &= y_tdt,\\
	dy_t &= \kappa(\eta-y_t)dt + \lambda\sqrt{y_t}dW_t .
	\end{align*}
	By conditioning on arrival rate (the hidden parameter) we determine the conditional likelihood function for VGSA
	\[ p(z_k|h^*) = \frac{2e^{\theta x_h/\sigma^2}}{\nu^{\frac{h^*}{\nu}}\sqrt{2\pi}\sigma\Gamma(\frac{h^*}{\nu})}\bigg(\frac{x^2_h}{2\sigma^2/\nu+\theta^2}\bigg)^{\frac{h^*}{2\nu}-\frac{1}{4}}K_{\frac{h^*}{\nu}-\frac{1}{2}}\bigg(\frac{1}{\sigma^2}\sqrt{x^2_h(2\sigma^2/\nu+\theta^2)}\bigg ) \]
	for a given arrival rate $dt^* = y_tdt$ and $h^* = y_th$ [39]. In order to give an intuition behind filtering, we follow the example outlined in [39]. We assume the hidden state (parameter set) evolves linearly, ie.
	\begin{align*}
	x_{t+1} &= ax_t + w_{t+1}, \\
	w_{t+1} &\sim N(0,\lambda^2),\qquad\text{f.s.\;} \lambda\in\Theta,
	\end{align*}
	where we assume $x_t$ (the prediction of the current state) is given along with the parameter set $\Theta$. Given observation $z_{t+1}$ at time $t+1$ we want to estimate $x_{t+1}$
	\[ \widehat{x}_{t+1} = \mathbb{E}(x_{t+1}|z_{t+1}). \]
	Assume the model price is given by $h(x_{t+1},\Theta)$ and assume the model price is related to the observation $z_{t+1}$ by
	\begin{align*}
	z_{t+1} &= h(x_{t+1},\Theta) + u_{t+1},\\
	u_{t+1} &\sim N(0,\sigma^2),\qquad\text{f.s.\;}\sigma\in\Theta.
	\end{align*}
	Both $\lambda,\sigma$ belong to set $\Theta$ so are already known. We now generate $M$ samples for $x_{t+1}$
	\[ x_{t+1}^{(i)} = ax_t + N(0,\lambda^2) \quad i=1:M,\]
	then generate $M$ samples for $u_{t+1}$
	\[ u_{t+1}^{(i)} = y_{t+1} - h(x_{t+1}^{(i)};\Theta), \quad i=1:M.\]
	The conditional likelihood function is
	\[ \mathcal{L}^{(i)} := \text{Likelihood}\Big(u_{t+1}^{(i)}\Big|u_{t+1}^{(i)}\Big) = \frac{\exp\Big(-\frac{(u_{t+1}^{(i)})^2}{2\sigma^2}\Big)}{\sqrt{2\pi}\sigma}, \]
	so the best estimate for $x_{t+1}$ is
	\[ \widehat{x}_{t+1} = \mathbb{E}(x_{t+1}|z_{t+1}) = \frac{\sum_{i=1}^{M}\mathcal{L}^{(i)}\times x_{t+1}^{(i)}}{\sum_{i=1}^{M}\mathcal{L}^{(i)}}. \]
	
	\subsection{VGSA Parameter Estimation via Particle Filtering}
	
	Using the integrated density for the VGSA process, we outline the particle filtering algorithm.
	\begin{enumerate}
		\item Initialize the states $x_0^{(i)}$ and weights $w_0^{(i)}$ for $i=1:N$ where $N$ is the number of price points available in the data. 
		\item Apply the extended Kalman filter to each state $x_k^{(i)}$ to obtain the transition update. Define $\widehat{x}^{(i)} = \kappa(\eta-x^{(i)})\Delta t$. The Gaussian approximation for the observation equation can be written as
		\[ z_k = h(x_k, B_k) = z_{k-1} + (\mu + \omega + \theta x_k)\Delta t + \sqrt{(\theta^2\nu + \sigma^2)x_k\Delta t}B_k. \]
		We then determine the Jacobians as follows:
		\begin{align*}
			A_{ij} &= (1-\kappa)\Delta t,\\
			W_{ij} &= \lambda\sqrt{x^{(i)}\Delta t},\\
			H_{ij} &= \theta\Delta t,\\
			U_{ij} &= \sqrt{\theta^2\nu + \sigma^2},
		\end{align*}
		The time update (prior estimate) is
		\[ \widehat{x}_{k|k-1} = x_k^{(i)} + \kappa(\eta + x_k^{(i)}\Delta t). \]
		The measurement update (posterior estimate) is
		\begin{align*}
			\widehat{x}_k &= \widehat{x}_{k|k-1} + K_k(z_k - h_{k|k-1}(\widehat{x}_{k|k-1})),\\
			&= \widehat{x}_{k|k-1} + K_k(z_k - (z_{k-1} + (\mu + \omega + \theta)\widehat{x}_{k|k-1})),
		\end{align*}
		which gives us the simulated state
		\[ \widetilde{x}_k^{(i)} = \widehat{x}_k^{(i)} + \sqrt{P_k^{(i)}}Z, \quad Z \sim N(0,1). \]
		\item We can now calculate the weights
		\[ w^{(i)}_k = w_{k-1}^{(i)} \frac{p(z_k|x_k^{(i)})p(x_k^{(i)}|x_{k-1}^{(i)})}{p(x_k^{(i)}|x_{k-1}^{(i)}, z_k)}, \]
		where $p(z_k|x_k^{(i)})$ is the integrated VGSA density defined earlier, $p(x_k^{(i)}|x_{k-1}^{(i)}) = N(x_{k-1}^{(i)} + \kappa(\eta - x_{k-1}^{(i)}\Delta t, \lambda\sqrt{x_{k-1}^{(i)}\Delta t )}$ and $q(x_k^{(i)}|x_{k-1}^{(i)}, z_k) = N(\widehat{x}_k^{(i)}, \sqrt{P_k^{(i)}}). $
		\item We then proceed through the rest of the particle filtering algorithm as explained earlier.
	\end{enumerate}
	The following VGSA particle filter algorithm will determine the log likelihood to be maximimized via some gradient descent optimization method:
	\lstset{tabsize=3, language=MatLab, numbers=left, numbersep=4pt, frame=single, commentstyle=\color[rgb]{0,0.6,0}}
	\begin{lstlisting}[mathescape]
	function [logl, estimates, errors] = VGSAParticleFilter(log_stock_prices,...
			 mu, N, kappa, eta, lambda, sigma, theta, nu)
	Nprices = N;
	Nsims = 100;     %Number of particles
	x = zeros(1,Nsims);
	xsim = zeros(1,Nsims);
	w = zeros(1,Nsims);
	u = zeros(1,Nsims);
	c = zeros(1,Nsims);
	dt = 1/252; %Trading days per year
	eps = 1e-10;    
	Pkk1 = zeros(1,Nsims); %a priori error ($P_{k|k-1}$)
	Pkk = zeros(1,Nsims); %a posteriori error ($P_{k|k}$)
	U = zeros(1,Nsims); %Gradient of h WRT measurement noise
	Kk = zeros(1,Nsims); %Kalman gain 
	W = zeros(1,Nsims); 
	xhat = zeros(1,Nsims);
	xk = zeros(1,Nsims);
	omega = log(1-theta*nu-sigma^2*nu/2)/nu;
	x0 = 1;
	P0 = 0.000001;
	x = x0 + sqrt(P0)*randn(1,Nsims);
	Pkk(:) = P0;
	A = 1-kappa*dt; %Jacobian WRT system process
	H = theta*dt; %Gradient of h WRT measurement noise
	logl = 0; %Log likelihood
	estimates = zeros(1,Nprices);
	errors = zeros(1,Nprices);
	wprev = ones(1,Nsims);
	for k=1:Nprices-1
		z = log_stock_prices(k+1)-log_stock_prices(k);
		xh = z-mu*dt-dt/nu*log(1-theta*nu-sigma^2*nu/2);
		x1_sum = 0;
		
		for i=2:Nsims
			%Simulate the state via extended Kalman filter
			%Time update
			xhat(i) = max(eps,x(i)+kappa*(eta-x(i))*dt); %prior transition update
			W(i) = lambda*sqrt(x(i)*dt); %Jacobian WRT system noise
			Pkk1(i) = A*Pkk(i)*A + W(i)^2; %prior error estimate 
			
			U(i) = sqrt(theta^2*nu+sigma^2)*sqrt(xhat(i)*dt);
			%Optimal gain
			Kk(i) = Pkk1(i)*H/( H*Pkk1(i)*H + U(i)*U(i));
			%Measurement update
			xk(i) = xhat(i) + Kk(i) * (z - (mu+omega+theta*xhat(i))*dt);
			Pkk(i) = (1.0-Kk(i)*H)*Pkk1(i); %Posterior error covariance matrix
			
			x1_sum = x1_sum + xhat(i);        
			
			%Simulate the state
			xsim(i) = max(xk(i) + sqrt(Pkk(i))*randn(1), eps);	
			%Calculate weights
			m = xk(i);
			s = sqrt(Pkk(i));
			%Normal density with mean m and stddev s
			q = 1/(s*sqrt(2*pi))*exp( -0.5*(xsim(i) - m)^2/(s^2));
			
			m = x(i-1) + kappa*(eta - x(i-1))*dt;
			s = lambda*sqrt(x(i-1) * dt);
			%Normal density
			px = 1/(s*sqrt(2*pi))*exp( -0.5*(xsim(i) - m)^2/(s^2));
			
			h = dt*xsim(i);
			%Arguments of Bessel function
			Kx = max(eps, 1.0/(sigma^2)*sqrt(xh^2*(2*sigma^2/nu+theta^2)));
			Knu = max(eps, (h/nu-0.5));
			
			%gammaln/besselk requires appropriate error handlers omitted here
			%VGSA integrated density
			pz = 2.0*exp(theta*xh/(sigma^2))...
			/(nu^(h/nu)*sqrt(2*pi)*sigma*gammaln(h/nu))...
			*(xh^2/(2*sigma^2/nu+theta^2))^(0.5*h/nu-0.25)...
			*besselk(Kx,Knu);
			%weights
			w(i) = wprev(i) * pz * px / max(q, eps);        
		end
		sumweights = sum(w);
		logl = logl + log(sumweights);		
		% estimates[i1+1] for z[i1] => error term
		estimates(k+1) = log_stock_prices(k+1)-(log_stock_prices(k) +...
			(mu+omega+theta*x1_sum/Nsims)*dt);
		errors(k) = (theta*theta*nu + sigma*sigma)*x1_sum/Nsims*dt;
		
		w = w./sumweights; %Normalize
		wprev = w;
		
		%Resample and reset weights
		c(1)=0;
		for i=2:Nsims
			c(i) = c(i-1) + w(i);
		end
		i = 1;
		for j=1:Nsims
			u(j) = 1.0/Nsims*(rand+j-1);
			while u(j) > c(i) && i < numel(c)
				i = i+1;
			end
			xsim(j) = x(i);
			w(j) = 1.0/Nsims;
		end
	end
	logl = -logl;
	end
	\end{lstlisting}
		%
		%
		%***********************************************************************************************
		%***********************************************************************************************
		%
		%
	\section{Optimal Parameter Set}
	\begin{adjustwidth}{2cm}{2cm}
				\textit{Whether cross-sectional option prices are consistent with the time-series properties of the underlying asset returns is probably the most fundamental of tests.}
	\end{adjustwidth}
	\hfill - D.S. Bates\nl
	The path shown in Fig. 6 will be used for backtesting the calibration and filtering procedures. It was generated with variance gamma using the following parameters: $\sigma = 0.28, \nu = 0.41, \theta = 0.1, T = 1, r = 0.1, q = 0$ over $N=2520$ corresponding to 10 price changes per trading day for one full year. We aim to fit VGSA to a VG path, here we describe the affect of the extra parameters of VGSA to gain some intuition behind path behavior.
	\begin{enumerate}[noitemsep]
		\item CIR time change process: $dy_t = \kappa(\eta-y_t)dt+\lambda\sqrt{y_t}dW_t$,
		\item $\kappa$: rate of mean reversion,
		\item $\eta$: long-term rate of change,
		\item $\lambda$: volatility of the time change.
	\end{enumerate}
	\begin{figure}[h!]
		\centering
		\includegraphics[scale=0.6]{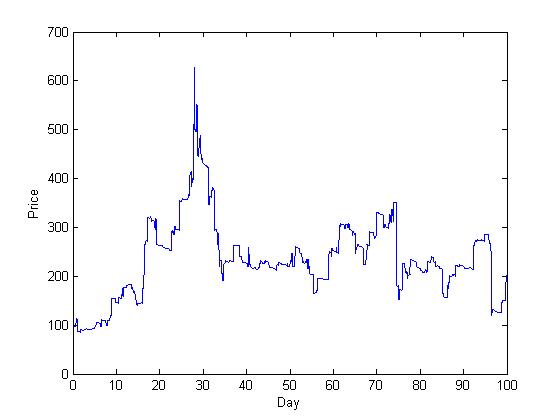}
		\caption{A typical path generated under VG law}
	\end{figure}
	For all optimization configurations, we impose non-negativity constraints on the above variables.
	\subsection{Backtesting RMEMC}
	We backtest the calibration procedure of section 5 by determining the set of option premiums associated with a stock following a VG process with arbitrary parameters then try to recover the original underlying process. Given random initial conditions satisfying nonegativity constraints for $\sigma, \nu$ we begin by calculating a series of option prices using the asset price at day 1. We use these estimated option prices and the actual observed option prices to determine the Kullback Leibler divergence between the estimated and actual price measures. Throughout the algorithm, at the end of each iteration the estimated pricing measure is used as the regularization prior for the next iteration. Each successive iteration generates the optimal pricing measure by adding the minimum amount of information to the optimal pricing measure for the previous day. Figure 7 shows the sum of squares option pricing errors (SSE) of two estimated pricing measures, with either high or low relative entropy (RE) with respect to the true distribution. The pricing errors are determined across a range of 30 option premiums calculated each trading day, calculated via Monte Carlo with both the true pricing measure and the estimated pricing measure.

	\begin{figure}[h!]
		\centering
		\includegraphics[scale=0.7]{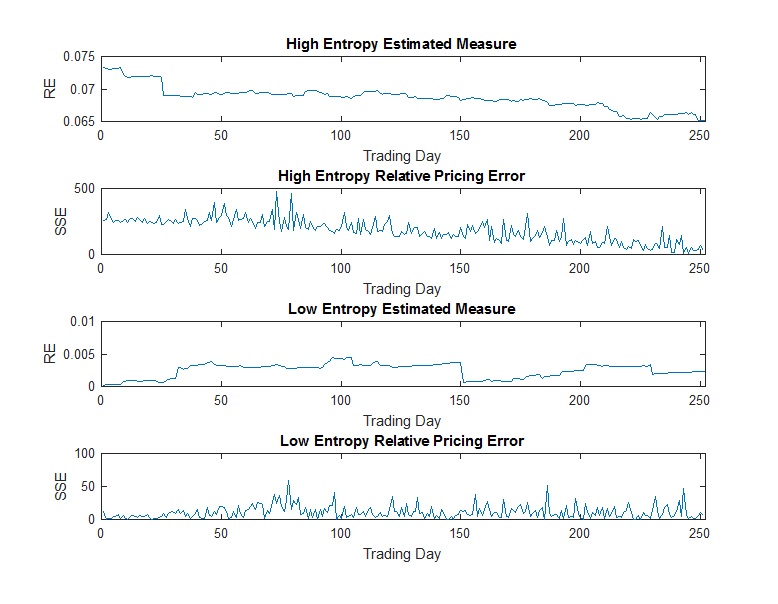}
		\caption{Performance of two predicted pricing measures where the relative entropy represents its 'distance' from the true distribution}
	\end{figure}
	Relative entropy is approximate since we used Monte Carlo simulation for VGSA. An advantage of using Monte Carlo to compute option prices as opposed to a transform or PDE method is that way can safely use out of the money options. While Monte Carlo does not suffer from out of the money price skewness, it requires high computational power to obtain a satisfactory level of accuracy.
	
	The minimal entropy martingale measure adds a degree of stability to the estimated pricing measure. Setting the initial condition at time $t$ equal to the optimal parameters of time $t-1$ generates the source of stability, but convergence to the true distribution will take longer depending on choice of regularization parameter. A sound question is whether to calculate the inter-day pricing measure using information gleaned from all time-series information of stock since time $t=0$. This is similar to a maximum likelihood procedure. Using only the optimal information from time $t-1$ we generate the next likely pricing measure as a sequence of approximating measures to the movement of the asset price. This sequence of approximating measures is Markovian.
	
	Beginning the calibration procedure with parallelized gradient descent algorithms iterating over a grid of initial points yielded the best outcome, similar to that of other global optimization schemes such as simulated annealing. How do we determine which initial condition has the lowest relative entropy with respect to the true distribution? The ability of the calibration procedure to converge to the true distribution is theoretically sound, but in practice it is very slow. In order to make substantial progress we need large stock price observations, which often is not available. As a stock price changes, so may its underlying distribution, and so we are effectively chasing the parameters of a dynamic distribution that's changing faster than the number of observations allow us to estimate. 
	
	Minimal entropy least squares calibration is very effective effective at resisting change in parameters however. If we manage to obtain the parameters of the true distribution, then for an appropriate choice of regularization parameter we can take new price movements into account while maintaining close proximity to the measure which accurately predicted historical prices. The notion of minimizing relative entropy is mathematically very similar to maximum likelihood. We see in the next section that non-Gaussian time series information can be accurately characterized via particle filtering, and so a joint method of particle filtering \textit{accounting} for option pricing information will likely produce much better results.
	
	\subsection{Backtesting the Particle Filter}
	We backtest the time series parameter estimation approach by first simulating via Monte Carlo a typical path of a stock following the VGSA process. We then try to recover the original parameters using our model of section 6.
	Where the nonzero value of $\lambda$ may account for the large swings of the estimate asset path corresponding to smaller swings of the true asset. Recall $\lambda$ is the volatility of the time change, the subordination process is what controls the mean reverting behavior of VGSA, and thus larger shifts in the subordinator may result in more intense price jumps. The filtering procedure appears to be very sensitive to volatility, as represented in Figure 8 which shows the true VG path along with the predicted paths.
	\begin{figure}[h!]
		\centering
		\includegraphics[scale=0.5]{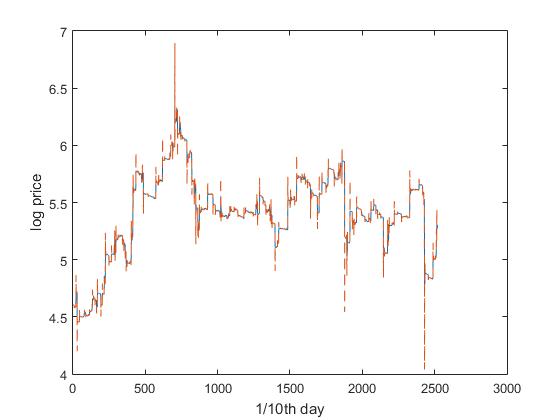}
		\caption{Estimated and True Asset Paths}
	\end{figure}
	
	\subsection{Which one is correct?}
	We have shown two different methods for extracting model parameters from market information. The cross sectional approach of least squares estimation of available option premiums should indeed agree with the parameters obtained from the time-series filtering approach. The filtering approach recovers the statistical parameter set from time-series information. The least squares calibration procedure recovers risk-neutral parameter set from cross-sectional information. 
	
	\begin{figure}[h!]
		\centering
		\includegraphics[scale=0.8]{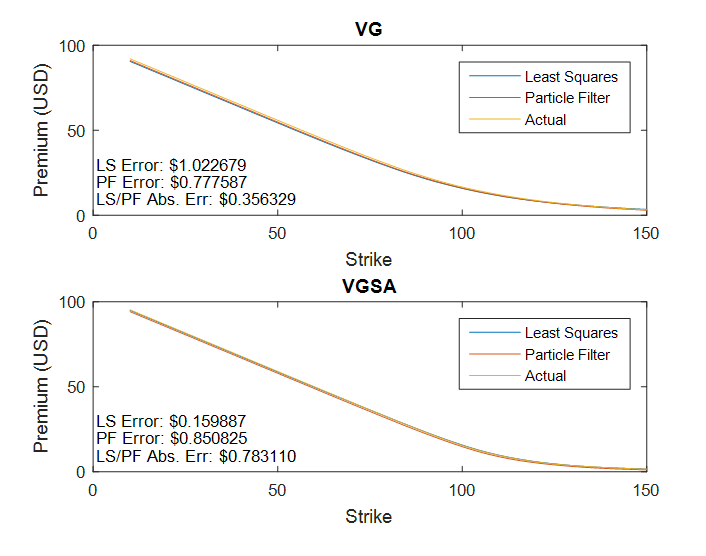}
		\caption{Least Squares/Particle Filter Estimated Option Premiums and Actual Premiums}
		\small
		\begin{flushleft}
		LS/PF Error: The absolute error between estimated and actual option premiums for a given strike. For each strike, 30 options were calculated, and so the error is absolute average.\\
		LS/PS Abs. Err.: The average absolute error between least squares and particle filter estimated option premiums. This is, on average, how much the RMEMC and PF algorithms agree.
				\end{flushleft}
	\end{figure}
	While it is meaningless to directly compare the optimal parameters from particle filtering and least squares calibration, we can measure the discrepancy between their estimated option premiums. Since RMEMC estimated option premiums do not have to be discounted but PF estimated option prices do, a simple modification to the Monte Carlo pricing algorithm can compare the estimated prices from both. We can see in Figure 8, that these algorithms do in fact agree on their estimated option premiums. Interestingly, the VG estimated prices from RMEMC and PF agree with each other more than they agree with the actual option prices. Much of this error is due to Monte Carlo. It is difficult to determine the effect on Monte Carlo pricing error on these estimates, it is however safe to assert that these methods do in fact agree with each other.

	%
	%***********************************************************************************************

	%***********************************************************************************************
	%
	%
	\section{References}
	{[1]} Abate, Joseph, and Ward Whitt. ``Numerical inversion of probability generating functions." Operations Research Letters 12.4 (1992): 245-251.\\
	{[2]} Alspach, Daniel L., and Harold W. Sorenson. ``Nonlinear Bayesian estimation using Gaussian sum approximations." Automatic Control, IEEE Transactions on 17.4 (1972): 439-448.\\
	{[3]} Andricopoulos, Ari D., et al. ``Extending quadrature methods to value multi-asset and complex path dependent options." Journal of Financial Economics 83.2 (2007): 471-499.\\
	{[4]} Andricopoulos, Ari D., et al. ``Universal option valuation using quadrature methods." Journal of Financial Economics 67.3 (2003): 447-471.\\
	{[5]} Arulampalam, M. Sanjeev, et al. ``A tutorial on particle filters for online nonlinear/non-Gaussian Bayesian tracking." Signal Processing, IEEE Transactions on 50.2 (2002): 174-188.\\
	{[6]} Avellaneda, Marco, et al. ``Calibrating volatility surfaces via relative-entropy minimization." Applied Mathematical Finance 4.1 (1997): 37-64.\\
	{[7]} Bates, David S. ``Maximum likelihood estimation of latent affine processes." Review of Financial Studies 19.3 (2006): 909-965.\\
	{[8]} Barndorff-Nielsen, Ole E., Thomas Mikosch, and Sidney I. Resnick, eds. Lévy processes: theory and applications. Springer Science \& Business Media, 2012.\\
	{[9]} Blake, Andrew, and Michael Isard. Active contours: the application of techniques from graphics, vision, control theory and statistics to visual tracking of shapes in motion. Springer Science \& Business Media, 2012.\\
	{[10]} Bollerslev, Tim. ``Generalized autoregressive conditional heteroskedasticity." Journal of econometrics 31.3 (1986): 307-327.\\
	{[11]} Bookstaber, Richard M., and James B. McDonald. ``A general distribution for describing security price returns." Journal of business (1987): 401-424.\\
	{[12]} Breeden, Douglas T., and Robert H. Litzenberger. ``Prices of state-contingent claims implicit in option prices." Journal of business (1978): 621-651.\\
	{[13]} Cappé, Olivier, Simon J. Godsill, and Eric Moulines. ``An overview of existing methods and recent advances in sequential Monte Carlo." Proceedings of the IEEE 95.5 (2007): 899-924.\\
	{[14]} Carr, Peter, and Dilip Madan. ``Option valuation using the fast Fourier transform." Journal of computational finance 2.4 (1999): 61-73.\\
	{[15]} Carr, Peter, and Dilip Madan. ``Saddlepoint methods for option pricing." Journal of Computational Finance 13.1 (2009): 49.\\
	{[16]} Carr, Peter, et al. ``Stochastic volatility for Lévy processes." Mathematical Finance 13.3 (2003): 345-382.\\
	{[17]} Chan, Terence. ``Pricing contingent claims on stocks driven by L\'{e}vy processes." Annals of Applied Probability (1999): 504-528.\\
	{[19]} Cont, Rama, and Peter Tankov. ``Calibration of jump-diffusion option pricing models: a robust non-parametric approach." (2002).\\
	{[20]} Cont, Rama, and Peter Tankov. ``Retrieving Lévy processes from option prices: Regularization of an ill-posed inverse problem." SIAM Journal on Control and Optimization 45.1 (2006): 1-25.\\
	{[21]} Cox, John. ``Notes on option pricing I: Constant elasticity of variance diffusions." Unpublished note, Stanford University, Graduate School of Business (1975).\\
	{[22]} Cox, John C., Jonathan E. Ingersoll Jr, and Stephen A. Ross. ``A theory of the term structure of interest rates." Econometrica: Journal of the Econometric Society (1985): 385-407.\\
	{[23]} Chan, Terence. ``Pricing contingent claims on stocks driven by L\'{e}vy processes." Annals of Applied Probability (1999): 504-528.\\
	{[24]} Csisz\'{a}r, Imre. ``I-divergence geometry of probability distributions and minimization problems." The Annals of Probability (1975): 146-158.\\
	{[25]} Derman, Emanuel, Iraj Kani, and Joseph Z. Zou. ``The local volatility surface: Unlocking the information in index option prices." Financial analysts journal 52.4 (1996): 25-36.\\
	{[26]} Dubner, Harvey, and Joseph Abate. ``Numerical inversion of Laplace transforms by relating them to the finite Fourier cosine transform." Journal of the ACM (JACM) 15.1 (1968): 115-123.\\
	{[27]} Dupire, Bruno. ``Pricing with a smile." Risk 7.1 (1994): 18-20.\\
	{[28]} Engl, Heinz Werner, Martin Hanke, and Andreas Neubauer. ``Regularization of inverse problems." Vol. 375. Springer Science \& Business Media, 1996.\\
	{[29]} Engle, Robert F. ``Autoregressive conditional heteroscedasticity with estimates of the variance of United Kingdom inflation." Econometrica: Journal of the Econometric Society (1982): 987-1007.\\
	{[30]} Esche, Felix, and Martin Schweizer. ``Minimal entropy preserves the Lévy property: how and why." Stochastic processes and their applications 115.2 (2005): 299-327.\\
	{[31]} Fama, Eugene F. ``The behavior of stock-market prices." The journal of Business 38.1 (1965): 34-105.\\
	{[32]} Fang, Fang, and Cornelis W. Oosterlee. ``A novel pricing method for European options based on Fourier-cosine series expansions." SIAM Journal on Scientific Computing 31.2 (2008): 826-848.\\
	{[33]} Fujiwara, Tsukasa, and Yoshio Miyahara. ``The minimal entropy martingale measures for geometric L\'{e}vy processes." Finance and Stochastics 7.4 (2003): 509-531.\\
	{[34]} Gordon, Neil J., Salmond, David J., and Adrian FM Smith. ``Novel approach to nonlinear/non-Gaussian Bayesian state estimation." Radar and Signal Processing, IEE Proceedings F. Vol. 140. No. 2. IET, 1993.\\
	{[35]} Gradshteyn, Izrail S. Table of integrals. Elsevier Science, 2014.\\
	{[36]} Handschin, J. E. ``Monte Carlo techniques for prediction and filtering of non-linear stochastic processes." Automatica 6.4 (1970): 555-563.\\
	{[37]} Harrison, J. Michael, and Stanley R. Pliska. ``Martingales and stochastic integrals in the theory of continuous trading." Stochastic processes and their applications 11.3 (1981): 215-260.\\
	{[38]} Heston, Steven L. ``A closed-form solution for options with stochastic volatility with applications to bond and currency options." Review of financial studies 6.2 (1993): 327-343.\\
	{[39]} Hirsa, Ali. Computational methods in finance. Crc Press, 2012.\\
	{[40]} Ito, Kazufumi, and Kaiqi Xiong. ``Gaussian filters for nonlinear filtering problems." Automatic Control, IEEE Transactions on 45.5 (2000): 910-927.\\
	{[41]} It\={o}, Kiyosi. On stochastic differential equations. Vol. 4. American Mathematical Soc., 1951.\\
	{[42]} Jackson, Nicholas, Endre Suli, and Sam Howison. ``Computation of deterministic volatility surfaces." (1998).\\
	{[43]} Jacod, Jean, and Albert Shiryaev. Limit theorems for stochastic processes. Vol. 288. Springer Science \& Business Media, 2013.\\
	{[44]} Javaheri, Alireza. Inside volatility arbitrage: the secrets of skewness. Vol. 317. John Wiley \& Sons, 2011.\\
	{[45]} Julier, Simon J., and Jeffrey K. Uhlmann. ``New extension of the Kalman filter to nonlinear systems." AeroSense'97. International Society for Optics and Photonics, 1997.\\
	{[46]} Kitagawa, Genshiro. ``Monte Carlo filter and smoother for non-Gaussian nonlinear state space models." Journal of computational and graphical statistics 5.1 (1996): 1-25.\\
	{[47]} Kulhav\'{y}, Rudolf. ``Recursive nonlinear estimation(a geometric approach)." Lecture Notes in Control and Information Sciences (1996).\\
	{[48]} Liu, Jun S., and Rong Chen. ``Blind deconvolution via sequential imputations." Journal of the american statistical association 90.430 (1995): 567-576.\\
	{[49]} Lord, Roger, et al. ``A fast and accurate FFT-based method for pricing early-exercise options under Lévy processes." SIAM Journal on Scientific Computing 30.4 (2008): 1678-1705.\\
	{[50]} Lord, Roger, and Kahl, Christian. ``Optimal Fourier inversion in semi-analytical option pricing." (2007): 2.\\
	{[51]} Madan, Dilip B., Carr, Peter P., and Eric C. Chang. "The variance gamma process and option pricing." European finance review 2.1 (1998): 79-105.\\
	{[52]} Madan, Dilip, and Marc Yor. ``CGMY and Meixner subordinators are absolutely continuous with respect to one sided stable subordinators." arXiv preprint math/0601173 (2006).\\
	{[53]} Madan, Dilip B., and Milne, Frank. ``Option Pricing With VG Martingale Components." Mathematical finance 1.4 (1991): 39-55.\\
	{[54]} Madan, D. B., and Seneta, E. ``Chebyshev polynomial approximations for characteristic function estimation: some theoretical supplements." Journal of the Royal Statistical Society. Series B (Methodological) (1989): 281-285.\\
	{[55]} Mandelbrot, Benoit B. ``New methods in statistical economics." Fractals and Scaling in Finance. Springer New York, 1997. 79-104.\\
	{[56]} Miyahara, Yoshio. ``Minimal entropy martingale measures of jump type price processes in incomplete assets markets." Asia-Pacific Financial Markets 6.2 (1999): 97-113.\\
	{[57]} Morozov, Vladimir Alekseevich. ``On the solution of functional equations by the method of regularization." Soviet Math. Dokl. Vol. 7. No. 1. 1966.\\
	{[58]} Morozov, Vladimir Alekseevich. ``The error principle in the solution of operational equations by the regularization method." USSR Computational Mathematics and Mathematical Physics 8.2 (1968): 63-87.\\
	{[59]} Nassar, Hiba. ``Regularized Calibration of Jump-Diffusion Option Pricing Models." (2010).\\
	{[60]} O'Sullivan, Conall. ``Path dependant option pricing under Lévy processes." EFA 2005 Moscow Meetings Paper. 2005.\\
	{[61]} Rubinstein, Mark. ``Implied binomial trees." The Journal of Finance 49.3 (1994): 771-818.\\
	{[62]} Ringn\'{e}r, Bengt. Prohorov's theorem. Centre for Mathematical Sciences. 2008\\
	{[63]} Rosiński, Jan. ``Series representations of L\'{e}vy processes from the perspective of point processes." L\'{e}vy processes. Birkhäuser Boston, 2001. 401-415.\\
	{[64]} Sato, Ken-Iti. L\'{e}vy processes and infinitely divisible distributions. Cambridge university press, 1999.\\
	{[65]} Samperi, Dominick. ``Calibrating a diffusion pricing model with uncertain volatility: regularization and stability." Mathematical Finance 12.1 (2002): 71-87.\\
	{[66]} Self, Steven G., and Kung-Yee Liang. ``Asymptotic properties of maximum likelihood estimators and likelihood ratio tests under nonstandard conditions." Journal of the American Statistical Association 82.398 (1987): 605-610.\\
	{[67]} Shreve, Steven E. Stochastic calculus for finance II: Continuous-time models. Vol. 11. Springer Science \& Business Media, 2004.\\
	{[68]} Tankov, Peter. L\'{e}vy processes in finance: inverse problems and dependence modelling. Diss. Ecole Polytechnique X, 2004.\\
	{[69]} Van Der Merwe, Rudolph, et al. ``The unscented particle filter." NIPS. Vol. 2000. 2000.\\

\end{document}